\documentclass[epj]{svjour}
\usepackage{graphics}
\usepackage[comma,square,numbers,sort&compress]{natbib}
\usepackage{hyperref}
\usepackage{lineno}
\usepackage{xspace}
\usepackage{xcolor}
\usepackage{epsfig}
\usepackage{lmodern}
\usepackage{slantsc}
\usepackage{relsize}
\usepackage{lipsum}% http://ctan.org/pkg/lipsum
\usepackage{multicol}% http://ctan.org/pkg/multicols
\usepackage{graphicx}% http://ctan.org/pkg/graphicx
 \usepackage{comment}
 \usepackage{enumitem}
\usepackage{adjustbox}
 \usepackage{makecell}
 \usepackage{multirow}

\def\babar{\mbox{\slshape B\kern-0.1em{\smaller A}\kern-0.1em B\kern-0.1em{\smaller A\kern-0.2em R}}}
\def\root{\mbox{\tt ROOT~}}
\def\github{GitHub~}
\def\gitlab{GitLab~}

\begin{document}
% 1) Help 
\newcommand{\todo}[1]{\textcolor{purple}{#1}}
\newcommand{\done}[1]{\textcolor{blue}{#1}}
\newcommand{\new}[1]{\textcolor{red}{#1}}
\newcommand{\old}[1]{\textcolor{orange}{#1}}
\newcommand{\change}[1]{\textcolor{olive}{#1}}
\newcommand{\ask}[1]{\textcolor{magenta}{#1}}
\title{Data Preservation in High Energy Physics}

\author{\textbf{DPHEP Collaboration: }
T.~Basaglia\inst{1}, 
M.~Bellis\inst{2}\thanks{\emph{Also at:} Siena College}
J.~Blomer\inst{1}\and
J.~Boyd\inst{1}\and
C.~Bozzi\inst{3}\and
D.~Britzger\inst{4}\and
S.~Campana\inst{1}\and
C.~Cartaro\inst{5}\and
G.~Chen\inst{6}\and
B.~Couturier\inst{1}\and
G.~David\inst{7}\thanks{\emph{Also at:} Stony Brook University}
C.~Diaconu\inst{8}\and
A.~Dobrin\inst{9}\and
D.~Duellmann\inst{1}\and
M.~Ebert\inst{10}\and
P.~Elmer\inst{11}\and
J.~Fernandes\inst{1}\and
L. Fields\inst{21}\and
P.~Fokianos\inst{1}\and
G.~Ganis\inst{1}\and
A.~Geiser\inst{12}\and
M.~Gheata\inst{9}\and
J.~B.~Gonzalez Lopez\inst{1}\and
T.~Hara\inst{13}\and
L.~Heinrich\inst{1}\and
M. Hildreth\inst{21}\and
K.~Herner\inst{14}\and
B.~Jayatilaka\inst{14}\and
M.~Kado\inst{1}\and
O.~Keeble\inst{1}\and
A.~Kohls\inst{1}\and
K.~Naim\inst{1}\and
C.~Lange\inst{20}\and
K.~Lassila-Perini\inst{15}\and
S.~Levonian\inst{12}\and
M.~Maggi\inst{22}\and
Z.~Marshall\inst{18}\and
P.~Mato Vila\inst{1}\and
A.~Me\v{c}ionis\inst{1}\and
A.~Morris\inst{17}\and
S.~Piano\inst{16}\and
M.~Potekhin\inst{7} \and
M.~Schr\"oder\inst{1}\and
U.~Schwickerath\inst{1}\and
E.~Sexton-Kennedy\inst{14}\and
T.~\v{S}imko\inst{1}\and
T.~Smith\inst{1}\and
D.~South\inst{12}\and
A.~Verbytskyi\inst{4}\and
M.~Vidal\inst{1}\and
A.~Vivace\inst{1}\and
L.~Wang\inst{6}\and
G.~Watt\inst{19}\and
T.~Wenaus\inst{7}
}

\institute{
%1
CERN, Geneva, Switzerland \and
%2
Cornell University, USA \and
%3 
INFN Ferrara, Italy\and
%4
Max-Planck-Institut für Physik, München, Germany \and
% 5 ----  
SLAC National Accelerator Laboratory, USA\and
%6 ----  
Institute of High Energy Physics, IHEP, CAS, Bejing, China\and
%7 ----  
Brookhaven National Laboratory, BNL, USA\and
%8 ----  
Aix Marseille Univ, CNRS/IN2P3, CPPM, Marseille, France \and
%9 ----  
Institute of Space Science, ISS,  Bucharest, Magurele, Romania\and
%10 ----  
HEP Research Computing, University of Victoria, BC, Canada\and
%11 ----  
Princeton University, USA\and
%12 ----  
Deutsches Elektronen Synchrotron, DESY, Hamburg, Germany\and
%13 ----  
High Energy Accelerator Research Organization, KEK, Tsukuba, Japan\and
%14 ----  
Fermi National Accelerator Laboratory, Batavia, USA\and
%15 ----  
Helsinki Institute of Physics, Finland\and
%16 ----  
INFN Trieste, Italy\and
%17 ----  
University of Bonn, Germany\and
%18 ----  
Lawrence Berkeley National Laboratory, Berkeley, USA\and
%19 ----  
IPPP, Durham University, Durham, UK\and
%20 ----  
Paul Scherrer Institut, Villigen, Switzerland\and
%21 ----  
University of Notre Dame, Notre Dame, USA \and
%22 ----  
INFN Bari, Italy
}

\date{Version published in Eur. Phys. J. C 83, 795 (2023) \\
DOI: \href{https://doi.org/10.1140/epjc/s10052-023-11885-1}{https://doi.org/10.1140/epjc/s10052-023-11885-1} \\ 
Received: April 18, 2023 / Accepted: July 20, 2022}
\abstract{
Data preservation is a mandatory specification for any present and future experimental facility and it is a cost-effective way of doing fundamental research by exploiting unique data sets in the light of the continuously increasing theoretical understanding. 
This document summarizes the status of data preservation in high energy physics. The paradigms and the methodological advances are discussed from a perspective of more than ten years of experience with a structured effort at international level. The status and the scientific return related to the preservation of data accumulated at large collider experiments are presented, together with an account of ongoing efforts to ensure long-term analysis capabilities for ongoing and future experiments. Transverse projects aimed at generic solutions, most of which are specifically inspired by open science and FAIR principles, are presented as well. A prospective and an action plan are also indicated.
}
%
%\PACS{
%      {PACS-key}{discribing text of that key}   \and
%      {PACS-key}{discribing text of that key}
%     } % end of PACS codes
%end of abstract

\vspace{2cm}

%https://www.overleaf.com/project/620130f8604b2abf8b2e8bfa
\maketitle
\pagebreak
\tableofcontents
\nopagebreak

\section{Overview} 

The issue of data preservation (DP) emerged with force as an important and community-wide issue in the field of high energy physics (HEP) at the end of the first decade of this century, as a consequence of the end of several large collider programs, such as HERA\footnote{The acronyms are described in the attached glossary.}, TeVatron, PEP-II  etc. By the end of the 2010's, the data preservation concept had been largely debated and formalised by an international working group, which rapidly was recognised by the International Committee for Future Accelerators\footnote{https://icfa.hep.net/} (ICFA) as an expert panel. Previous pioneering preservation initiatives with large and complex data sets (e.g. the JADE or LEP experiments) were also included. 

The working group produced a number of recommendations\cite{DPHEPStudyGroup:2009gfj}: (i) urgent action had to be taken to organise the long term data preservation at the experiment and site levels, with identified resources, (ii) global approach was necessary towards an international collaboration and (iii) careful consideration of new, advanced technologies to address the issue of data preservation was needed. Most notably, a significant advance was made in defining the concept of ``data preservation'', that includes in fact all aspects related to a productive data analysis activity: digital data, metadata, publications, software, databases, documentation etc. A key issue was identified to be the organisation of data analysis activities in the long term, in particular by transforming and adapting the collaborations (new rules for governance, lighter procedures, flexible membership etc.), as well as considering to open the data for reuse by enlarged communities. 

The ICFA panel, with a strong support from major laboratories and in particular CERN\footnote{A symposium held at CERN in 2009 paved the way to a global collaboration \url{https://indico.cern.ch/event/70436/}.} initiated an international collaboration called Data Preservation in High Energy Physics  (DPHEP) with the primary goal to foster the international collaboration and mutual support across HEP collaborations in enriching the scientific return of HEP data. DPHEP issues regular reports \cite{DPHEPStudyGroup:2012dsv,DPHEP:2015npg}.

The present document presents an overview of the DP activities worldwide~\footnote{This document reflects the contributions presented at the ``Third DPHEP Collaboration Meeting'' \url{https://indico.cern.ch/event/1043155/timetable/}.}. The status of the DP activities in the participating HEP laboratories are briefly summarized below :
\begin{itemize}

    \item{\bf DESY} (Deutsches Elektronen Synchrotron, Hamburg) has been the host of PETRA electron-positron collider, where JADE experiment collected data until 1983. The JADE data resurrection and preservation~\cite{Bethke:2022cfc}, a particularly instructive and pioneering DP project is now hosted at the Max Plack institute for Physics (\textbf{MPP}) in Munich, where a multi-experiment framework is explored as well. \\ The main DP activity in DESY at present is related to the former HERA collider experiments H1 and ZEUS, that collected data until 2007. They adopted different preservation philosophies: H1 (migration and encapsulation) and ZEUS (encapsulation). Their data analysis systems and collaborations are functional. Successful transitions to the DP systems are reported. The publications continue with a regular rate and the publication plan for the next years includes a dozen potential papers. The objective is to keep systems running and available through 2030. New institutes are joining the collaborations, in synergy with the future Electron-Ion Collider (EIC) experiments. 
%\item{\bf MPP} in Munich has played a pioneering role and provides expertise on data preservation, starting with the particularly instructive process of JADE data resurrection~\cite{Bethke:2022cfc}. A multi-experiment framework is explored (JADE, HERA, OPAL). Recent activities include ``JADE on a desktop'', a project dedicated to making JADE data as portable as possible. 
\item{\bf CERN} (European Organisation for Nuclear Research, Geneva) hosted the LEP, the largest electron-positron ($e^+e^-$) collider to date. LEP provided an unique data set at highest energies, that may become instructive in the preparation of the future circular colliders (FCC-$ee$)~\cite{Blondel:2021ema}. A low rate but clearly identified LEP data and software activity is reported, with refreshed standards and technologies resulting from open data and open science initiatives created and developed on site. Moreover, the LHC (Large Hadron Collider) activity is in full swing at CERN and the corresponding data preservation issues are treated in strong connection with open data approaches.  \\ CERN is the host laboratory of DPHEP, maintains the DPHEP portal and ensures the operational management, which is essential for the collaboration. A rich panel of transverse projects have been developed at CERN towards an open usage of data and analysis, primarily from LHC experiments, but with a significant potential to incorporate data from other experiments. 

\item{ \bf SLAC } The \babar\ experiment at the $e^+e^-$ PEP-II collider, hosted at the Stanford National Accelerator Laboratory (SLAC), collected high precision data for the heavy flavour studies until 2008. The collaboration has been active in pursuing data preservation and producing a constant and significant publications flow, with the 600$^{th}$ paper published recently~\cite{BaBar:2022ahi}. However, SLAC support ended in February 2021. The data is copied to GridKa for user analyses. The analysis support is also provided at the University of Victoria, that hosts the collaboration\footnote{\url{https://babar.heprc.uvic.ca/}} as well. For recovery purpose, the 1.7 PB of data is also copied to CERN %(1.2 PB done and 0.5 PB ongoing)
as well as hosted at CC-IN2P3. A user infrastructure for ongoing analyses and documentation is hosted by the HEP Research Computing (HEP-RC) group at the University of Victoria, Canada. 
\item{\bf{KEK}} The Belle I experiment data, collected until 2010 at KEKB collider, situated at High Energy Accelerator Research Organization (KEK, Tsukuba). The Belle I data preservation has been pursued in parallel with the construction of Belle II experiment. As a result, the transition and the overlap between Belle I/II experiments is the main feature of DP activities at KEK. Belle I data is readable in the Belle II framework. The objective is to maintain Belle I data through 2023, at which point the precision will be exceeded by the new data.

\item{\bf FNAL} The TeVatron is an world unique proton--anti-proton collider, situated at the Fermilan National Laboratory (FNAL, Chicago). The TeVatron hosted the experiments CDF and D0 until the stop of the data taking in 2011. A transition to a DP system for both CDF and D0 took place in 2012. Data is stored/saved at FNAL and in Italy, but there is no intention to maintain the analysis facility. One can note the 500th D0 paper in 2020~\cite{D0:2020ujb}, as well as the recent $W$ mass measurement from CDF~\cite{CDF:2022hxs}, which are also illustrative of the interest for a long term preservation of unique data sets. MINERvA, a neutrino-nucleon scattering experiment, also started a DP project.
\item{\bf BNL} The PHENIX experiment at the heavy ion collider RHIC, hosted by the Brookhaven National Laboratory (BNL), stopped data taking in 2016. PHENIX pursues a data and analysis preservation program. There are further reports on contributions to DP activities at LHC, together with a newly started reflection on DP at the future electron-ion collider (EIC).%, discussed with NPC.
\item{\bf IHEP} The BES III experiment at the institute for high energy physics (IHEP) in Beijing was expected to stop data taking by 2022, but a prolongation to 2030 is envisaged. A sustained activity on DP has been pursued along with the data taking, with the objective is to preserve the data for at least 15 years. %Strong support for national and international DP activities has been expressed.

\end{itemize}

\begin{figure*}[hhh]
\centering
\includegraphics[width = 0.99\textwidth]{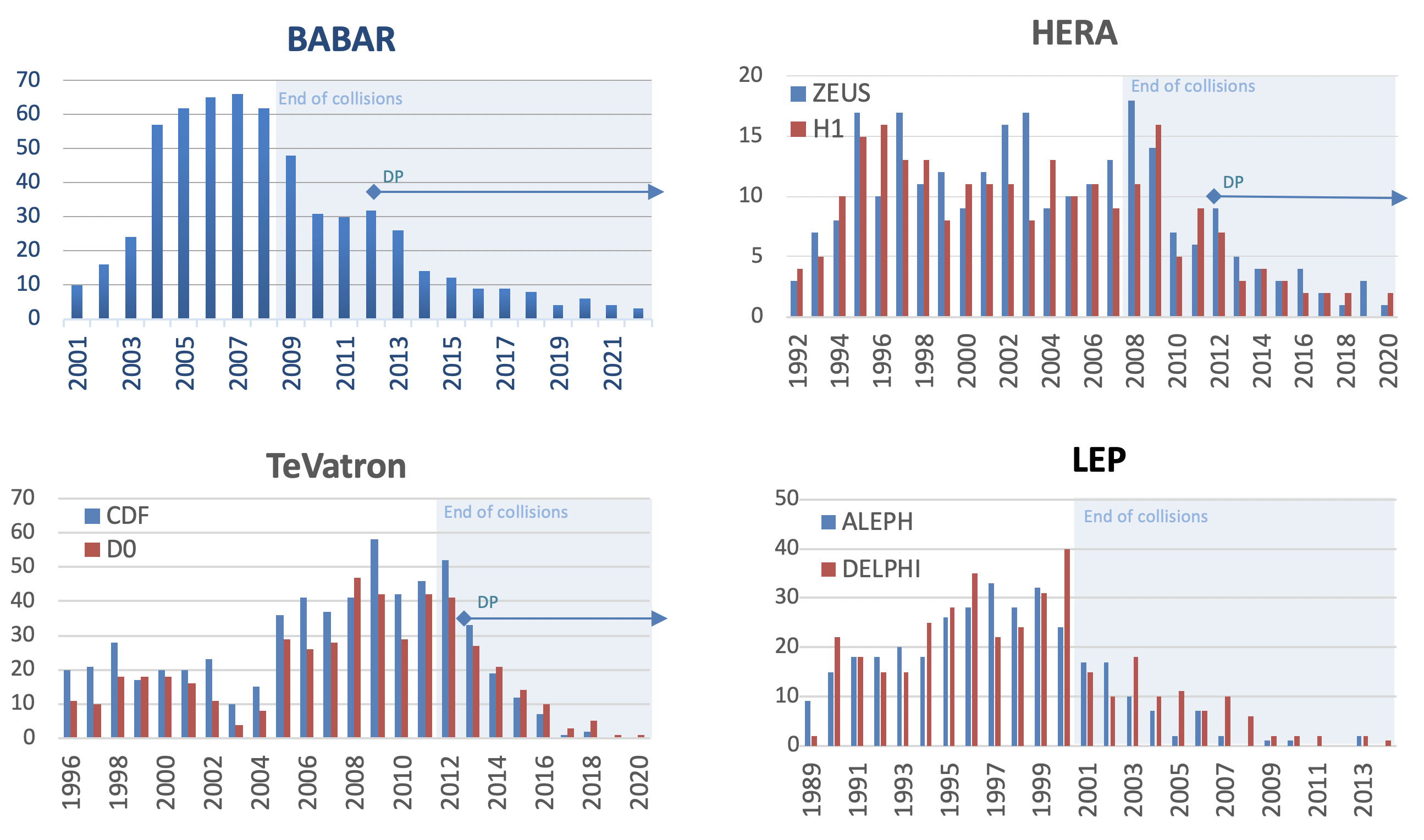}
\caption{
The publication record for four major experimental facilities. The period after the data taking is indicated by the shaded area. The coverage by a \textit{dedicated} data preservation project ( i.e. pursued in addition to the regular computing activities) is also shown as an arrow labelled DP. 
}
\label{fig:2022PROD}
\end{figure*}

Figure \ref{fig:2022PROD} illustrates the scientific production at major experimental facilities as a function of time and illustrates the significant research outcome obtained after the end of the data taking. 
%One can estimate the fraction of publications attributed to dedicated data preservation system to about 10\%.\footnote{It should be noted that those publications cover only the usage by the collaborations themselves. The subsequent usage of those publications also adds to the scientific impact of the preserved data, but it is not taken into account here.}

Based on the above input from the participating experiments and laboratories, the global status of data preservation in HEP can be summarized as follows:
\begin{itemize}
 \item The implementation of DPHEP recommendations led to an enhanced scientific outcome in HEP. There are tangible effects of dedicated projects using the DPHEP recommendations (level classification and robust choices, formalisation of long term collaborations needs for data stewardship, new technological solutions such as virtualisation etc.). 
  \item The concepts of DPHEP have been used in extending the scope of data preservation towards open science; for instance the CERN Open Data policy~\cite{CERN-OPEN-2020-013} has adopted the DPHEP classification. Moreover, the updates of the computing models at the LHC~\cite{LHC_COMPUTING_UPDATES_2014}  incorporate data preservation as a specification, thereby increasing the chances for a smooth and productive transition into ``preserved mode'' at the end of the program.  The community roadmap issued by the HEP Software Foundation covered the DP aspects as well~\cite{HEPSoftwareFoundation:2017ggl}.
 \item The investments/outcome balance is close to the estimates from 2012: about 10\% of the overall publications have been obtained using a dedicated data preservation project (as opposed to a tolerated prolongation of a slowly freezing computing system), for which the global costs did not exceed a few per mille of the total experiments costs. This remarkable balance demonstrates the previous prediction~\cite{DPHEPStudyGroup:2009gfj} that a proactive data preservation action supports research at low cost~\footnote{It should be noted that those publications cover only the usage by the collaborations themselves. The subsequent usage of those publications also adds to the scientific impact of the preserved data, but it is not taken into account here.}. 
\item Some data sets are still in danger or have evolved towards an unusable state. For instance, the \babar\ data set has been decommissioned from the host laboratory and is transferred to other computing centres. It can be noted that the DPHEP community has been active in searching for solutions. It is likely that the situation will repeat with other data sets in the future. 
\item The survey demonstrates that future/new experiments are likely to manifest an interest for the preserved data sets, for training, testing or even addressing new subjects. The emblematic JADE/LEP example~\cite{Bethke:2009ehn}, where the re-analysis of JADE data lead to an adequate precision towards a common analysis with the more recent LEP data, is now being reproduced for HERA and EIC. Indeed, some EIC groups joined HERA collaborations (still active and organised around well-preserved data sets) in order to test new paradigms and train
 for the future experimental program. It is likely that the preparation for FCC-$ee$ will benefit from the preserved LEP data sets, at least at the level of data model design and analysis frameworks preparation. 
 \item The proof that complicated data analysis frameworks can be preserved with significant gains may stimulate more communities and experiments to join the reflection and implement suitable strategies. An example is offered in this document by MINERvA, a neutrino-nucleon scattering experiment at Fermilab.
 %Dark Matter, Fixed target at CERN and JLAB; Neutrinos baseline project (Dune, Ks) etc. 
 \item New types of analysis methods (e.g. machine learning) and data access (e.g. zenodo\cite{https://doi.org/10.25495/7gxk-rd71}) are being tested on preserved data sets, thereby leading to new results or methods for long term preservation and open access.
 \item Open science policies implemented for recent or ongoing data sets are essential for the long term robustness of the data preservation. Moreover, the open science paradigms are being implemented for the already preserved (older) data sets. This approach requires nevertheless a significant effort, that could be supported in the context of a demonstrated scientific interest. 
\item It should be noted, however, that the preservation systems remain fragile. The better understanding of the main weaknesses and an improving and proactive attitude towards DP did not remove the danger of a catastrophic loss. The key issue is the person power, that functions for most of the examples of preserved data sets in a voluntary mode. The sharp technological steps are particularly dangerous (for instance disappearance of 32-bit platforms) and need permanent attention and dedicated action.

\end{itemize}
The overall conclusion is that the existence of an international structure such as DPHEP, oriented to a long term perspective for data preservation, minimally supported and hosted by a large laboratory (CERN), has served successfully as reference to well-defined projects with shorter time scale, that have obtained clear advances. 

%\newpage

\section{What is data preservation?}
It may be useful to recall the scope and the goals of data preservation in HEP~\cite{DPHEPStudyGroup:2012dsv}. 

 \paragraph{What is data?}: the short answer is ``everything that was created as a result of planning, running and exploiting an experiment''. Indeed, there is a persistent confusion associating ``data'' to an operating system files, i.e. bits on a memory support such as disks, tapes, etc. This simplistic approach is not operative for running experiments and cannot be used to plan for a functional long term data preservation. Although the file system is of course essential, it is by far insufficient to perform a data analysis. Instead, we propose to define the ``data'' in HEP -- but also in any experimental work using digital/computing systems -- as the multiverse of all specific inputs, outputs and digital tools used by a group of researchers in order to obtain a novel result: 
\begin{itemize}
    \item digital data files: raw and processed, control/configuration, meta-data, environmental parameters, operational data, databases, etc.,
    \item software in all its forms (front-end, trigger, middleware, reconstruction, classification include machine learning setups, high-level analysis, visualisation etc.),
    \item documentation files (internal/public notes, publications, manuals, contracts, photographs, technical drawings), and
    \item organisation and diffuse knowledge files: rules and procedures, contracts, minutes, meetings and slides, news, blogs, logbooks, address books, outreach material etc.

\end{itemize}
All these aspects have been considered and synthetically encapsulated in the so-called data preservation levels~\cite{DPHEPStudyGroup:2012dsv}, briefly described in the next chapter. It is worth noting that some of those components may not exist in a digital format or may not be sufficiently structured during the data taking, and therefore need to be recreated in a robust format by the preservation project. 

 \paragraph{What is preservation?}: the process of transforming a ``high intensity / rapidly changing'' computing system into a ``low intensity / slowly evolving'' computing system while conserving the capacity of extracting new science from the ``data'' (within its definition of above). ``Preservation'' in this context is \textit{not} a freezer, nor a herbarium, a museum, an album, etc. It is, as the concrete examples presented in this document demonstrate, \textit{a sustained and technologically demanding operation}. In other words, the preservation is a complex project, that has to take into account the data typology, the research goals, the available resources and the collaboration decisions. It may involve choices in data aspects presented above and therefore deliberate data losses/dismissal, as well as significant restructuring and new processing. Furthermore, 
 %it cannot be a simple ``playground with real data'', 
 the system should be usable to reproduce already-published results but also to explore new ideas. Therefore, the system has to preserve at least a part of (and ideally all) the unexplored knowledge of the accumulated experimental data (see the discussion on DP levels below). Under those conditions, the DP system has to tackle important tasks, such as:
 \begin{itemize}
     \item ensure physical existence of data from a digital point of view (see data definition above, all this has to be physically saved and secured at long term -- and that includes software, of course) -- note that this is the simplest and basically solved aspect of DP in HEP.  
     \item as an obvious (and relatively easy to solve) aspect of the previous item: identify and provide computing and storage resources.
     \item ensure the functionality of the whole system, identify the potential risks and take appropriate measures as technology and community evolve. The level of complexity differs for the various aspects of the data. The simplest examples include the digital files, the documentation etc. that need only storage and access, i.e. rather standard operations independent of the experiment complexity in general. In contrast, specific experimental software and databases are much more difficult to keep functional across technological changes (hardware, operating systems etc.). 
     \item ensure unambiguous and permanent data validation~\cite{Ozerov:2013isa} and results reproducibility, naturally setting the ground for data reuse and open access\cite{Chen:2018drk}. 
    \item define and identify the human resources related  to the research plan.\footnote{This aspect is particularly interesting since the bulk of the long term activities are done on a voluntary basis, so they escape the usual needs-resources dialog with the funding agencies. This aspect is addressed also in the perspective of a cost-benefits analysis in section~\ref{costs-benefits}}
     \item oversee and manage the collaborative work and manage the preserved data analysis activity according to the DP design. 
    \item define and implement data access policies, i.e. for which purpose and under which formal regulations the data can be used, including opening the access to data to new collaborators and/or releasing the data to larger (not pre-identified) communities.
     \item observe and update the physics case of the preserved data. It should be noted that the technical solutions and the necessary choices on the information to be dismissed while designing a long term preservation system should decouple as much as possible from the epoch-related physics case. Indeed, the door should remain open for unexpected analyses.
 \end{itemize}
 The goals of a data preservation system as expressed in~\cite{DPHEPStudyGroup:2012dsv} intrinsically comply with what has come to be known as FAIR principles~\cite{FAIR:2016}.  Indeed, the data has to be easy to find (F) and accessible (A), and therefore -- in a HEP collaborative context -- (re)usable (R). The interoperability (I), identified as one of the long term goals ten years ago, is becoming a built-in specification of the recent computing systems as well. Concrete steps have been achieved, with a few examples given in section~\ref{DPsystems}, with a strong incentive originating from the open science policy or within structural projects such as WLCG. However, a clear strategy for a FAIR approach over the entire HEP field (including past, present and future experiments) is still to be defined. 

The data preservation process implies a careful inventory of the existing "data archipelago", re-mapping its components in order to improve the navigation with the smallest possible effort in the long term. Important decisions have to be made, in particular on what should be preserved in line with the scientific objectives, the available resources and the collaboration perspectives\footnote{The DP process can be summarized as  "Consider everything, prepare as much as you can, preserve what is possible".}. A structured approach for addressing those crucial aspects is described in the next section. 

\section{Methodologies of HEP data preservation}

This section summarizes the generic model of data preservation now used in many experiments to refer to the chosen long term data preservation  models. 

\subsection{Data Preservation models and frameworks}

\subsubsection{Preservation levels}

Four generic DPHEP data preservation levels have been defined, supporting different use cases and expected efforts. The levels are organized in increasing complexity and are reported in Table~\ref{tab:levels}.

\begin{table*}[htb]
    \centering
%    \begin{tabular}{|c|l|l|}
    \begin{tabular}{|c|p{7.5cm}|p{8cm}|}
    \hline
    Level &  Model   &  Use Case \\
    \hline
    \hline
    1 & Provide additional information & Publication-related information search \\
    \hline
    2 & Preserve the data in simplified form & Outreach, simple training analysis \\
    \hline
    3 & Preserve the analysis-level software and data format & Full scientific analysis based on existing reconstruction \\
    \hline
    4 & Preserve the reconstruction and simulation software and raw data & Full potential of the experimental data \\
    \hline
    \end{tabular}
    \caption{Definition of the preservation models, in order of complexity.}
    \label{tab:levels}
\end{table*}

Guided by those generic levels, the experiments can choose the use case they intend to support and adapt the preservation model to the corresponding level. Higher levels include lower levels, so that if level 4 is chosen all the uses cases covered by levels 1-3 are also covered.

The first level involves providing information in addition to the published result with the purpose of improving the understanding of the result and/or the ability to use the result for a further high-level analysis. The additional information may include the exact values used of a published plot, extra tables, data hidden in assumptions use to derive the published, or result meta-data related to the running conditions, for example. A challenge associated with this level is the technology choice for storing the additional information. Global information infrastructures, such as the ones used by the running experiment or others used by the community, such as INSPIRE, may be beneficial for the robust preservation process. Care has to be taken to make sure that the technology chosen is kept alive or can be migrated to keep access to the additional data persistent. Usually solutions allowing export to text files in ASCII format have the highest grade of preservation.

The second level consists of preserving the data in a different format, such that it can be read by non-experiment specific software. Typically only some information is kept, for example the run information (integrated luminosity, collision energy, etc.) and the 4-vectors of the reconstructed event particles, the total energy and so on. The format can be as simple as a CSV table in ASCII, for a maximum portability, or some common binary format, such as \root data structures (TTrees), which have proven to be readable over at least two decades. The format for level 2 preservation is typically not enough for a new analysis; it is typically used for outreach and educational purposes.  This and the previous level are useful for reinterpretation analyses based on published synthetic data (such as data points, simplified n-tuples, likelyhood functions etc.). Moreover, one should be noted that the basic formats considered to be standard and stable at a given time may still evolve in the longer term.

The third level consists of preserving the experiment data in the original data format used for analysis as well as the software required to read and process those data. This is typically sufficient to perform a complete analysis if the related reconstruction and detector calibration are adequate for the purpose. The requirements on the experiment software preservation are much stronger than in lower levels. They can be mitigated by using the existing software to create intermediate objects which can be read and, for example, visualized with recent software.

The fourth level consists of preserving the data and the software in a format that the full chain is available, for example a new reconstruction to include new calibration constants, or improved algorithms. This requires also the possibility to generate new sets of simulated events with new and/or improved versions of the generator codes. This level is the most demanding in terms of software preservation, requiring the preservation of the whole environment, including many dependencies. The benefits are evident, retaining full flexibility for future use.

\subsubsection{Frameworks}
% {\it Reminder of concepts (levels), frameworks (freeze, virtual, modernize, suitcase, call-911….)}

Preserving data means preserving the physical support of the data themselves (the bits) and the ability to use them, as described above. Bit preservation consists of making sure that the data are stored in a safe mode on reliable and usable supports, which is usually achieved by employing certification procedures. 
%Certification processes to ensure that the bits are in a "healthy" state. 
Such certification procedures, developed by the space science community~\cite{certiso1, certiso2}, have also been considered for HEP experiments~\cite{jshiersNaples}. These protocols require that the data can be read by test programs run regularly and that they are kept on reliable and functional hardware. Data centers such as the one at CERN copy the data to new storage (tapes) every 18 months, which allows for sanity checks, for example making sure that all the files are still available, and to keep up-to-date with storage technology.

For a HEP experiment, ensuring the ability to use the data is more than just being able to read them back and write them out to a new medium. For the data to remain useful all the activities typical of an HEP experiment (generation of signals and backgrounds, simulation, reconstruction, analysis, etc.) must remain possible. This means that the experiment software ecosystem must continue to run. Because the experiments commonly  use customized software, keeping these ecosystems alive is an important challenge that HEP collaborations face.

There are essentially two ways to achieve this: freeze the last validated experiment software and create the conditions to continue to run the frozen software; or continuously port the software to the latest stable operating system. 

Freezing the software requires keeping the possibility to have access to the original operating system. This is usually achieved via virtual machines or keeping alive nodes still running the original operating system (these can be laptops with a bunch of disks attached -- the so-called "suitcase model" or "freezer"). This solution has some serious drawbacks; for example, the evolution of security requirements might break the connection of these machines to the rest of the world.\footnote{Problems of this kind are not only academic: SLC5-based CMS open-data analysis stop working when the data servers raised the TLS requirements to levels that the SLC5 clients could not cope with~\cite{gganisNaples}.}

% (Porting requires validation.)
Porting the software means adapting the existing software to a new operating system environment. A crucial part of porting is validation, which is making sure that the software behaves as expected. Porting the experiment software to a new OS is part of the experiment lifetime activities, so one could expect that the process is already established and should not present many difficulties. However, even for the most automated of the experiments, the validation of a new OS environment relies on the feedback of the physics groups, which is something fading out as soon as the experiment goes into preservation mode. A generic validation framework was discussed in~\cite{DPHEPStudyGroup:2012dsv} and implemented in several examples of DP systems presented in this document.

\subsection{Supervision models}
One of the key ingredients of the data management plans for long term preservation includes the status of the collaboration that steers the experiment. 
It is interesting to note that the organisation of experimental collaborations depends not only on the size and complexity of the experimental device, but also on their status in the data taking process. 

Various stages of organisation can be defined:

{\bf 0: Organisation during experiment proposal.} The collaboration structures itself such that the plans for a new experiment are pursued. The role of the physics case working group is critical in the first steps, with the experimental aspects being progressively enforced. From R\&D, production and global construction, the technical groups evolve during the data taking into running, maintenance and performance groups. The collaborations already produce significant amount of data (simulations, data bases, plans etc.), while the actual data and documentation design process is still ongoing. 

{\bf 1: Organisation during data taking.} The organisation during the data taking is focused on the running efficiency (experimental performance, data quality, data preparation and availability) and on extracting the full scientific information according to the publication plans. The data and the simiulations are  re-processed regularly and frequently. The community is fully focused on the scientific output. As for the previous stage, the data preservation was historically ignored during the data taking. However, more recent experiments do take into account those aspects and have developed dedicated task forces and adopted public policies for data preservation and access. This organisational form is perfectly illustrated at present by the LHC experiments.

{\bf 2: Organisation after data taking.} The end of data taking induces the need to adapt both immediately and longer after the data taking, during analysis and collaboration funding times. A strong decrease in the global common and institute-level funds is observed. The technical personnel is rapidly moved to other projects. A pressure to ``move on'' is generally manifested from the laboratories and funding agencies. In that context, not less, but more organisation is needed. The publication plans have to be fully consolidated with person power commitments (the reliability of which has to be carefully evaluated, to avoid overoptimistic schedules). The usual competition across the collaboration to obtain highlight results is not a management asset anymore (no multiple groups on the same subject, reduced ability to trigger a high-intensity initiative such as a conference rush or an urgent data processing, etc.). Each subject is basically covered by a group (or a single person) and a list of open subjects is normally emerging as well -- the latter naturally provides a clear and concrete argument for data preservation\footnote{The list of open subjects (i.e. not covered by person power commitments) tends to decrease slower than expected or even to grow after the end of the data taking for at least two reasons: lost of person power and new subjects emerging.} .  In this context, the collaborations usually re-organise for a more flexible, though rigorous, frame. This stage can be done in two sub-steps, depending on the available work-force and support from the participating institutes and the host laboratory.

{\bf 3: Organisation after the collaboration funding scheme.} Usually, the collaborations exist officially during the funding periods as agreed by the Memoranda of Understanding. After this period, the institutional bodies (funding agencies, institutional boards, etc.) diminish or stop their involvement.  Experience shows, however, that the scientific collaborations may extend well after the end of the official funding (HERA and \babar, for instance). In that case, the collaboration model of governance usually changes, giving a more prominent role to the active collaborators. This form of organization strongly relies on the support from the host laboratory, since the data stewardship has to be ensured by the host laboratory computing centre, usually as a continuation and with limited but sufficient technical support. For the distributed computing case (LHC, for instance), the funding agencies and the corresponding computing centers responsible for data hosting need to continue some support for the long term data stewardship in a coordinated way.

{\bf 4: Rescue organisational scheme.} This organisation scheme is to be activated when:
\begin{itemize}
\item the host laboratory stops support and announce no long-term commitment. 
\item the official collaboration/data stewardship is stopped with no further plans (no step 3 is clearly defined).
\end{itemize}
Examples include LEP (a clear step 3 has not been defined, although elements of technical support and continued publications process are present) and \babar\ (data support stopped at SLAC). The collaborations have found solutions by the initiative of individuals and proactive groups that continue to access and steward data sets (LEP at CERN, JADE at MPP Munich) or by moving data sets to other facilities and preserving the physics case (\babar\ at GridKa/Germany, CC-IN2P3/France and HEP-RC Victoria/Canada). No universal recipe can be applied here, but some actions still have to be taken in the form of a preservation project. Indeed, if/when the above listed conditions are encountered, taking no action necessarily implies decommissioning (deleting) the data.\footnote{The host laboratories and collaborations should be warned that any rescue operation vaguely imagined for a later stage is nearly impossible: storing and freezing the files and the latest version of the software is certainly not a substitute for a preservation project.}

These variants show that during the life of a collaboration, the organisation plays a fundamental role in the data longevity. While a few collaborations have succeeded in defining strong collaboration models that persisted for a long time (3), it is likely that, in the longer term, the ``data children'' will have to live their lives (4). In that case, the ambition of a full preservation model may need to be revised to a more simplified configuration that preserves an essential part of the information, mostly for historical and cultural purposes. In this context, the matter of standardisation and common format is very relevant (see for instance section \ref{key4hep_section}).

The data preservation issue used not to be discussed at all in the initial steps of the experiment (0 and 1), since it was most commonly considered as an end-of-the-run operation, which is far from optimal for the long term. This has changed in the past decade, with data preservation becoming one of the specifications of the experimental design. An illustrative example, although not referring to brand new experiments, is the update of the LHC experiments computing models for Run 2~\cite{LHC_COMPUTING_UPDATES_2014}. The preparation of future experiments at FCC and EIC, as well as the community roadmap~\cite{HEPSoftwareFoundation:2017ggl} confirm this trend as well. Moreover, in parallel and very often as a common and synergistic effort with data preservation, initiatives for opening data for larger communities and outreach have strongly emerged in the past years. Those initiatives also impact the internal organisation and in turn offer a higher reliability for the long term data preservation. 
\begin{comment}

The main point being the transitions and the corresponding data survival probability.

Peter: ``At some point you have to let your ``data children" live their lives ... So experiments could perhaps build in a clear ``sunset" expectation, along with something like the ``level 2.5" with simplified software dependencies."

DP Policy at the host labs (poetry)

 when during the collaboration lifecycle, which/how much expertise is required? (e.g. when can one define a common data format -- experts still there, performance constraints already relaxed) 
\end{comment}
\subsection{Preservation and openness}

%\new{Tibor, Lukas}

Open data, open software, and open science principles help to facilitate long-term data and knowledge preservation from several different points of view. 

Firstly, opening data may lead to simple ``lots of copies keep stuff safe'' usage scenarios. This increases confidence against data loss and helps to ensure data survival especially for smaller data set size scenarios. However, opening data and keeping them accessible for truly long term has inherent maintenance costs. The open data platforms may have to ensure the data integrity against ``bit rot'' but also need to be involved with future data format conversions to prevent obsolescence. The truly long-term data preservation and availability may therefore require guarantees by the open data providers to ensure long-term access to preserved data. The open data policies of experimental collaborations may have to take carefully into account the long-term data availability for periods extending beyond the experiment funding lifetime. It is interesting to explore connections with data management policies of hosting laboratories.

Secondly, opening data provides an opportunity to make it more robust. Opening data for larger communities beyond the context of the original experiment involves deepening data curation and data stewardship processes with the aim of making data more understandable, not only to a new generation of researchers, but also to non-specialists. A good example may be the rising importance of collaborations between particle physicists and the machine learning community. The data must be opened and described in a way that is understandable to other communities. The interoperability can be facilitated by richer data semantics. The robustness of the data-opening process is further assisted by carefully documenting the data provenance and the data validation principles.

Thirdly, opening data facilitates making preserved data more ``actionable''. The ultimate goal of data preservation is to facilitate future data reuse. The FAIR data stewardship principles advocate making data Findable (i.e. discoverable by both humans and machines); Accessible (i.e. available and obtainable by common protocols); Interoperable (i.e. syntactically and semantically understandable to a wide community of users) and Reusable (i.e. sufficiently described and shared for future reuse). The opening of preserved data make them ``actionable'' by simply lowering the barrier to providing data validation scripts or data analysis operational examples, the periodical execution of which helps to ensure the FAIR-ness of data as a function of time. Data preservation and openness thus often go hand in hand in facilitating future data reuse beyond the original data acquisition and analysis contexts.

\subsection{Funding and valuing data preservation}
\label{costs-benefits}
\begin{table*}[thb]
    \centering
    \begin{adjustbox}{width=0.75\textwidth}
    \begin{tabular}{|c||p{0.8in}|p{1.4in}|p{1.0in}|p{1.1in}|}
    \hline
         {\bf Program } & {\bf Data taking stopped }& {\bf Publications before 2012 } & {\bf Publications after 2012 } & {\bf Ratio in \%}  \newline Scientific output increase \\  \hline
        \babar & 2008 & 471 & 154 & 33\% \\     \hline
        H1+ZEUS & 2007 & 436 & 62 & 14\% \\     \hline
    \end{tabular}
    \end{adjustbox}
    \caption{Number of publications of HERA and \babar~ programs before and after the switch to dedicated data preservation systems. The reference year to define the long-term scientific production is taken as 2012, corresponding roughly to five years after the end of the data taking. }
    \label{tab:scientificreturn}
\end{table*}

The specific support for data preservation has different sources:
\begin{enumerate}[label=C\arabic*.]
    \item Host laboratories allocate person power and computing resources.
    \item Collaborating laboratories participate in the effort: replicate or take over data and computing systems and provide technical assistance.
    \item Researchers and engineers participate outside their main research area.
    \item Innovative computing projects, including pluri-disciplinary open science initiatives, may offer attractive opportunities for data preservation and are therefore an indirect source of support.
    \item The proximity of a follow-up experiment clearly helps in structuring and supporting a data preservation project.
\end{enumerate}
Given the success of most of the data preservation projects registered with DPHEP and presented below, the issue of allocated resources is a very interesting one. 
The most successful experiments have benefited from explicit host laboratory support in the initial phase (C1). This extra support allowed a definition of a specific project, for which the investments can be accounted for as ``data preservation costs''. According to the previsions from DPHEP initial documents and in agreement with the few projects observed in the past years, the direct investments in dedicated DP projects correspond to $\mathcal{O}(10)$ FTE-years with a very marginal investment in material\footnote{The costs for maintaining older data sets within hots laboratories computing centers beyond the costs foreseen in the MoU can be considered as negligible, since those data sets became relatively small and could be considered as a common "social security" service.} The C1 item can be compared with the total experimental costs that are, for the kind of collaborations considered here (HERA, \babar\ etc.) of a few $\mathcal{O}(10^{3})$ FTE-years (plus the constructions costs, usually corresponding to multi-hundred millions). Within this perspective, one can very approximately estimate that the investment in a DP project corresponds to at most a few per mille from the total cost of the experiment.

Those costs are to be compared to potential benefits:
\begin{enumerate}[label=B\arabic*.]
    \item New publications -- counting here those executed with a strong involvement of the dedicated DP systems.
    \item Publications made by other groups/people using the new publications produced at B1. 
    \item Preserving the scientific expertise and the leadership in the field of the experiment, possibly boosting the transition to a new experiment
    \item Technology expertise in robust data preservation. Improved ability to plan for new experiments and preserve their scientific potential at long term. 
\end{enumerate}

The listed items cannot all be straightforwardly quantified. An estimate of the publication enhancement rate is illustrated in Table~\ref{tab:scientificreturn} for two programs where dedicated data preservation systems were installed for the long term phase.

Therefore, the cost/benefits balance can only be counted in a very simplistic way, for instance with the ratio C1/B1. Normalizing this ratio to the entire experiment, it comes out from the few exemplary cases cited below that the cost-benefit ratio of the preserved data are very favourable, since investments of the order of a few per mille lead to about 10\% gains in the publication record. 

Going beyond this quick estimate, which offers nevertheless a very encouraging indication of the positive balance in favour of the preservation of data, a refined analysis of the benefits and costs would be very interesting in order to understand the overall landscape. This further analysis, which is not within the scope of the present document, would have to answer a number of interesting questions:

%\new{CERN/CAP peprojects hople (Alex Kohls?) / Achim / (ask more people)/ \Daniel}

\begin{enumerate}

\item Why did the interest in the data not stop after the data taking? The ``common sense'', expressed by some prominent scientists at the end of the data taking, was to ``publish your last paper and leave''. This approach assumes that the initial publication plan is exhaustive and once accomplished, the resources should be relocated to new projects, which is a classical objective-resources managerial approach. Still, a small but motivated community voluntarily kept data alive for many years and extracted unique science from it, well beyond the ``local n-tuples'' philosophy that eventually perpetuates only very specialised analyses. 

\item How are the human resources accounted for by the funding agencies or labs? Is doing analysis on preserved data subversive, tolerated or highly valued? 

\item How are the publications valued in the ``long-term'' analysis mode of a collaboration? What is the impact of those publications? Are the authors able to claim visibility and recognition?

\item How is the value of this (new) science displayed?  What is the full cost (and who is supporting it) to promote this 10\% of additional science? 

\item How is measured the value of data for outreach and education? 

\item How is HEP data contributing to the human culture as a whole (like in arts, e.g. a painting, or a piace of music, which cannot be valued just in terms of investment, resources and financial transactions)?

\item What global resources were used 5 and 10 years past the end of the experiment to keep systems alive and publish?

\item Are the DP requirements compatible with the running experiments conditions? How much extra investments are needed to make "fresh" data suitable for a long term preservation and how those investments can be optimised further when considering open data and open science aspects?

\item How are future projects supporting, stimulating and shaping data preservation projects and how are the cost and benefits of this transfer of knowledge accounted for?

\end{enumerate}
All these questions and many more, when considered in the perspective of the scientific outcome that continues to come from some experiments more than 15 years after data taking stops, indicate a ``data preservation miracle'', where science continues to be extracted at low cost from preserved data sets for which the access and the complexity are not obstacles for a small but highly motivated community.

\section{Experiment reports}
\begin{table*}[thb]
    \centering
    \renewcommand{\arraystretch}{1.9}
    \begin{adjustbox}{width=\textwidth}
    \begin{tabular}{|c|c|c|c|c|c|c|} 
    \hline
\makecell{{\bf Laboratory}/\\{\bf Collider}} & {\bf Experiment} & \makecell{{\bf Data taking}\\ {\bf period }}& \makecell{{\bf Preservation}\\{\bf Level}} & {\bf Data Volume}	& {\bf Present status} & \makecell{{\bf Collaboration} \\ {\bf supervision model} } \\  \hline \hline
%PETRA/JADE
\makecell{DESY/PETRA} &JADE   & 1979--1986 & 4 &  1~TB  &   \makecell{Analysis running on preserved data; \\ migrated from DESY to MPP }   & 4 \\		\hline			
% LEP
\makecell{CERN/LEP} & \makecell{ALEPH, DELPHI,\\ L3, OPAL}	& 1989-2000 & 4 & 0.5 PB  &   Analysis running on preserved data & 4 \\		\hline		
% HERA
\multirow{2}{*}{DESY/HERA}&H1  & \multirow{2}{*}{1992 -- 2007} & 4 & 0.5 PB	& \multirow{2}{*}{Analysis running on preserved data} & \multirow{2}{*}{3} \\ %\cline{2-6}
                            & ZEUS  &  & 3/4  & 0.2 PB &  & \\ \hline
% BABAR
\makecell{SLAC/PEP II}& \babar\  & 1999--2008 & 4 &	2 PB &	 \makecell{Analysis running on preserved data; \\ migrated from home lab to different centers} & 4 \\ \hline
%BELLE I	
\makecell{KEK/KEKB}&Belle I & 1999-2010 & 4 & 4 PB	& \makecell{Analysis running on preserved data; \\ Compatible with Belle II computing}  & 2 \\ \hline
%Tevatron	
\multirow{2}{*}{FNAL/TeVatron}&DØ  &  \multirow{2}{*}{1983--2011}	& 4  & 8.5 PB  & \multirow{2}{*}{Archived on tapes}& \multirow{2}{*}{4}\\ %\cline{2-6}
                               &CDF &            & 4 & 9 PB	 &  & \\ \hline
BNL/RHIC & PHENIX &	2000--2016  &  3 &  25 PB   & Analysis running on preserved data  & 3  \\		\hline
FNAL/$\nu$-beam & Minerva  &  2010--2019	& 3 &   10~TB  & Analysis running  & 2\\		\hline
IHEP/BEPCII  &BESIII	& 2009--2030	& 4 & 6 PB & Collecting and analyzing data & 1 \\ \hline
CERN/LHC & \makecell{ALICE, ATLAS,\\CMS, LHCb}	  & 2010-2040  & 4  & $\mathcal{O}\mathrm{(1EB)}$   &  Collecting and analyzing data &1 \\		\hline
    \end{tabular}

    \end{adjustbox}
    \caption{The data preservation status for the experiments presented in this document.}
    \label{tab:DPHEP_EXPERIMENTS}
\end{table*}

This section contains reports from a number of high-energy collider experiments connected to DPHEP. The experiments are presented in the order of their end of data taking, therefore starting with the longest preserved data set of JADE experiment and ending with running experiments at the LHC. The main features and the present status of the preserved data sets discussed in this report are presented in table~\ref{tab:DPHEP_EXPERIMENTS}.

 \subsection{JADE}
 \label{experiment:JADE}
 
 The JADE experiment was one of the experiments located at the PETRA $e^{+}e^{-}$ storage ring at DESY in Hamburg, Germany~\cite{Bethke:2022cfc}. %The JADE detector comprised accurate tracking, fast multi-hit electronics, measurement and identification of photons, electrons and muons.  A schematic view of the JADE detector is given in Fig.~\ref{fig:JADE:detector} and a more detailed description can be found in Ref.~\cite{Naroska:1986si}
 %
\begin{comment}

\begin{figure*}[tphb]
\centering
\includegraphics[width = 1.0\textwidth]{figures/JADE/JADEDet.jpeg}
\caption{Longtitudual cross-section of JADE detector. The diameter of the Jet Chamber is about 1~m.}
\label{fig:JADE:detector}
\end{figure*}
\end{comment}
%
 The experiment took data between 1979 and 1986 in the center-of-mass range between $12$ and $46.6$~GeV. The results from the JADE experiment were published within a regular collaboration structure between 1979 and 1991. Important scientific results of the JADE collaboration are the (co-)discovery of the gluon, the establishment of jet-physics and tests of Quantum-Chromodynamics (QCD). Other highlights are studies of the  hadronisation process via the string effect, electro-weak precision tests, two-photon physics and searches for then not confirmed particles of the Standard Model, i.e. the top quark and Higgs boson, and searches for New Physics like Super-Symmetry, free quarks etc.
%\textbf{Preserved JADE data}

The data includes the collision events recorded by the JADE detector  at energies between $12$ and $46.6$ GeV and the Monte-Carlo (MC) simulated events. Most of the preserved MC simulated event samples were generated in the 2000s, during the initial resurrection of the JADE software. These samples were produced with then contemporary MC event generators Herwig6~\cite{Corcella:2000bw}, Pythia6~\cite{Sjostrand:2006za}, Jetset~\cite{Sjostrand:1993yb}, with parameter settings as used by the OPAL experiment. Now in the 2020s these samples can be superseded with modern and more precise MC simulations. Both the data and MC simulated events are 
also preserved in a processed form of \root~\cite{Antcheva:2011zz} or HBOOK/PAW~\cite{Brun:1988pg} analysis frameworks $n$-tuples, which are suitable for many QCD-related analyses and are compatible with similar n-tuples of the OPAL experiment, see Sec.~\ref{LEP:OPAL}.

The preserved data includes the calibration information and the luminosity tables. The integrated luminosities at the main energy points range from 1.46/pb at 14~GeV to 150/pb at $33.8-36$~GeV. This corresponds to O(1000) hadronic events at the low energy points and about 35.000 hadronic events at $33.8-36$~GeV. 
As of 2020s, the JADE data with the total size just below 1~Tb can be considered as 'tiny'. With such a small data size, it makes little sense even to discuss the costs/resources that should be dedicated to the physical storage of data. However, for the convenience of access to the available data, the data are stored (and password/certificate protected) in local discs in MPCDF\footnote{Max-Planck Computing and Data Facility, Garching, Germany} archive file system, in the MPCDF dCache storage system and in the MPCDF ownCloud cloud storage. The diversity of storage instances provides an excellent opportunity for the access and analysis of the data from the Grid (dCache~\cite{Behrmann:2008zza} instance), or from an local desktops using ownCloud\footnote{\url{https://owncloud.com/}}. The data consists of about 35.000 files with a total size of 806~Gb.

%\textbf{Preserved JADE software}
The preserved JADE software consists of the original codes designed to process the JADE data, some MC event generators from the 1980-2000, the detector simulation routines for the JADE experiment, some calibration codes, the event display, and the analysis routines which were used to
create the ROOT/PAW n-tuples mentioned above. In addition to that some interface codes to modern MC event records were added. 

The original JADE software has evolved on many different platforms and after the first resurrection~\cite{MovillaFernandez:2002zz} in the middle of the 2000s consisted of approximately 50.000 lines of Fortran code running on IBM AIX4.3 systems with the IBM Fortran runtime. The environment the JADE software required for compilation and execution required also GNU or AIX binutils, C runtime and CERNLIB, see Sec.\ref{DPTECH:CERNLIB}. The JADE event display required a specific graphics package HIGZ~\cite{Bock:1988pd}. 

The main goal of the next update of JADE software was not only to port it from the old IBM AIX operating system to a modern Linux environment, but also to assure its portability and eliminate the need of any complex and/or exotic environment requirements. Therefore, JADE build system based on make was replaced with the {\sc CMake} build system. The codes were updated to be compatible with modern GNU Fortran and several other compilers. Not all of the used Fortran compilers are free, and not all Fortran runtimes have the support of the 
essential features for the JADE software (e.g.\  mixed endianess I/O), therefore only the GNU Fortran and Intel Fortran toolchains and runtimes are practically useable.

Thanks to the portability of the {\sc CMake} build system it became possible to compile the JADE codebase not only on Linux but also for the first time on MacOSX. The  dependence of the original codes on the CERNLIB and HIGZ was undesirable, so to avoid these dependencies the required CERNLIB and HIGZ functions were emulated with a help of the \root analysis framework, with an optional  dependence on  CERNLIB. With the performed code updates it became possible to create new MC generated event samples using modern MC event generators, and to re-reconstruct the original JADE data. Some functionality of the event display was also restored. The computing model of the JADE data re-analysis would include the processing of the data (and/or MC simulated events) into plain \root n-tuples and subsequent 
steps performed using the \root framework.

The JADE codes were publicly accessible from the dedicated JADE site for a long time\footnote{\url{ https://www.mpp.mpg.de/en/research/data-preservation/jade}}. After the updates, the codes were put in a public repository account in \github\footnote{ \url{https://github.com/andriish/JADE}}.
The usage of \github has allowed for regular automated builds of the JADE software on modern Linux and MacOSX platforms. All the dependencies needed for the software are available from open source projects.

%\textbf{JADE documentation and cultural heritage}
The physics papers of JADE are available on INSPIRE. Many of those are scanned copies of the hard copies from CERN or KEK
%\footnote{High Energy Accelerator Research Organization, 1-1 Oho, Tsukuba, Ibaraki, Japan} 
libraries. The JADE web site and the \github site mentioned above include as well: the full list of JADE physics papers, technical notes, 
scanned logbooks, data and software preservation documentation.
In addition, the technical notes are available as hard copies at MPP in Munich.

In addition to the equipment, also the data and the software of a physics experiment 
can be considered as part of the history of physics and of physicists. A part of the archive of photographs from the times of the active collaboration was made public 
on the JADE web site at MPP in Munich.

%The physics papers of JADE are available on InSpire. Many of those are scanned copies of the hard copies from CERN or KEK\footnote{High Energy Accelerator Research Organization, 1-1 Oho, Tsukuba, Ibaraki, Japan} libraries. The full list of JADE physics papers is available at \url{ https://www.mpp.mpg.de/en/research/data-preservation/jade}. Technical notes are available as scanned copies at \url{ https://www.mpp.mpg.de/en/research/data-preservation/jade}, in GitHub as well as hard copies at MPP. The logbooks are available as scanned copies and are available at \url{ https://www.mpp.mpg.de/en/research/data-preservation/jade} and in GitHub. The data and software preservation documentation is available at \url{ https://www.mpp.mpg.de/en/research/data-preservation/jade} and in GitHub.

%\textbf{JADE cultural heritage}

%In addition to the equipment, also the data and the software of a physics experiment  can be considered as part of the history of physics and of physicists. A part of the archive of photographs from the times of the active collaboration was made public at \url{ https://www.mpp.mpg.de/en/research/data-preservation/jade}.

%\textbf{Most recent JADE analyses}

Although many $e^+e^-$ experiments were conducted after the end of data taking period of JADE, the energy range covered by JADE remains unique and could only be accessed in future experiments, e.g.\ in just a few days of data taking at FCC-$ee$. Therefore, the JADE data was a source of important Quantum Chromo-Dynamics (QCD) studies in the last decades. As an example, in the most recent JADE analyses~\cite{Schieck:2012mp,Pahl:2009zwz} a relatively competitive extraction of the QCD strong coupling constant $\alpha_S(M_Z)$ was performed using these data.

%\textbf{JADE preservation policy}

As of 2022 MPP offers support for any possible re-analysis efforts of the JADE data under supervision of the JADE members. We recommend that the results are published according to the collaboration guidelines discussed in 2009. The JADE group at MPP is eager to join the most modern developments of the Data Preservation in the context of the JADE Data preservation and therefore contributing to the CERN OpenData initiative is under active discussion.

In March 2022 the JADE collaboration approved making the JADE data public and the decision was backed by the DESY Directorate. Therefore anyone can re-analyze the data. Any results obtained with JADE data and/or software should be supplemented with a note
``
We thank the JADE collaboration and DESY for making the data and corresponding software publicly available. The data analysis presented here has not been reviewed by these entities and is the sole responsibility of the authors.''.

%\new{Workshop contributions + invited;}
%\new{Editors: the speakers; we can invite more contributions (is there something missing? JLAB?)}
\subsection{ALEPH, DELPHI, L3, OPAL}

The Large Elecron-Positron ($e^+e^-$) (LEP) collider at CERN, installed in a dedicated 27~km circular tunnel that is nowadays re-used for the Large Hadron Collider (LHC), operated four experiments between 1989 and 2000, at energies up to 204 GeV. 
LEP remains to date the most powerful lepton collider, and its followup at higher energies, the Future Circular Collider (FCC-${ee}$) will not produce data before the decade 2040, providing an example of the enormous gap in time that can occur between successive frontier research programs. 

\begin{figure*}[hhh]
\centering
\includegraphics[width = 0.6\textwidth]{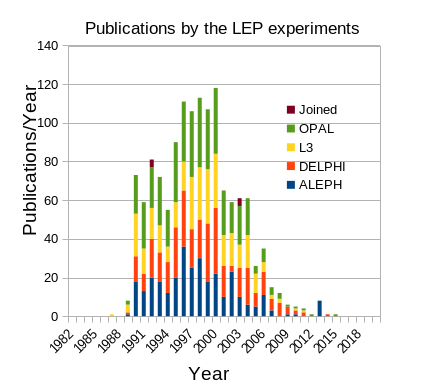}
\caption{LEP publications per experiment. Note that those publications only reflect the work done by the collaborations themselves. Further usage of those publications, also enhancing the impact of the preserved data, is not accounted for by this figure. }
\label{fig:LEPpubl}
\end{figure*}

The LEP  experiments initially stored their data sets in the local storage system CASTOR at CERN with two copies on tape. As part of archive service modernisation, the tape-resident data is about to be migrated to  CTA, the successor of CASTOR. Already in 2015, the frequently used data types have been copied to the EOS systems, making it available to users without the need to stage any tapes. The preservation and resurrection of the software library CERNLIB played an important role and is described in Sec.~\ref{chap:cernlib}. Access to data on tape and disk is regulated by the respective experiment policies and is generally restricted. Fig.~\ref{fig:LEPpubl} shows the number of publications by the four LEP experiments over time. It shows a long tail of continued activity after the end of data taking in the year 2000. A short status report for each of the four experiments is given below.

\begin{comment}
    
\begin{itemize}
    \item Do we want to make a statement about FCC/LEP synergies (also FCC chapter)?
    \item mention FCC-ee schedule as an factor defining LEP data uniqueness?
    \item Reference the joint technical note outlining a roadmap for a consolidated effort in preserving LEP data exploiting above synergies and target timelines?
    \item Add a clearer statement on expected lifetime of last working platforms
    \begin{itemize}
        \item Separate timelines for main sw and graphical (event displays)?
        \item Carefully stated risk statement and alternatives.
        \item Crucial role of validation in case of  alternatives need be deployed.
    \end{itemize}
    \item Scanned/OCRed documentation in CDS (== current hostlab document service) is seen as common goal (Aleph?)
    \begin{itemize}
        \item Detect data quality issue also on documentation side
    \end{itemize}
    \item Delphi/Opal  are using a common CERNVM image
    \item Publication plots [Delphi/Aleph] (see Fig.~\ref{fig:LEPpubl})
\end{itemize}
\end{comment}

\subsubsection{ALEPH}

The ALEPH libraries to access the collected data and to produce simulated data were compiled with g77 for the 32-bit x86 processors and are still functional on platforms supporting/emulating this architecture. The most recent validated operating system is SLC6, which is available under CVMFS and can be used, for instance, with Singularity. The ALEPH software is also available in CVMFS and \gitlab at CERN, and rely heavily on the 32-bit/g77 version of  CERNLIB. To move to a more recent architecture will require the re-compilation of the entire software followed by an in-depth validation campain. Part of the measured data and simulated data produced by the ALEPH experiment were made available in a simplified format, which is accessible in EOS with much reduced dependency on the original software. This approach promises to be a longer-term solution to access the existing datasets translated in a more recent HEP data format. In case of need of new simulated data with upgraded or new models the full software stack implementing the entire ALEPH workflow is still necessary.

\subsubsection{DELPHI}

The DELPHI experiment has moved its software stack to modern technologies, e.g. \gitlab at CERN for the sources and CVMFS for the binaries, while archiving older binaries on EOS. During 2022 the full stack was ported to 64-bit as support for 32-bit libraries is vanishing. This was possible thanks to the efforts to revive CERNLIB, see chapter~\ref{chap:cernlib}. The experiment software heavily relies on CERNLIB thus a validated and complete 64-bit version of CERNLIB is a pre-requisit for the future of the DELPHI software stack. The revised DELPHI software stack supports data analysis frameworks, simulation, reconstruction and event visualisation. Builds are available for the following operating systems: CentOS7, CentOS-Stream 8 and 9 and Alma 8 and 9 in both 32~bit and 64~bit, and Ubuntu 18, 20 and 22 (64~bit only). A special challenge was the removal of a dependency on a commercial software package in the event display. 
When logging into CERN interactive services with a DELPHI registered account, the login scripts will automatically select the correct version of the stack and set all environment variables as needed. It is also possible to initialise the stack manually by simply sourcing a script on CVMFS.

\begin{figure*}[htb]
\begin{center}
\epsfig{figure=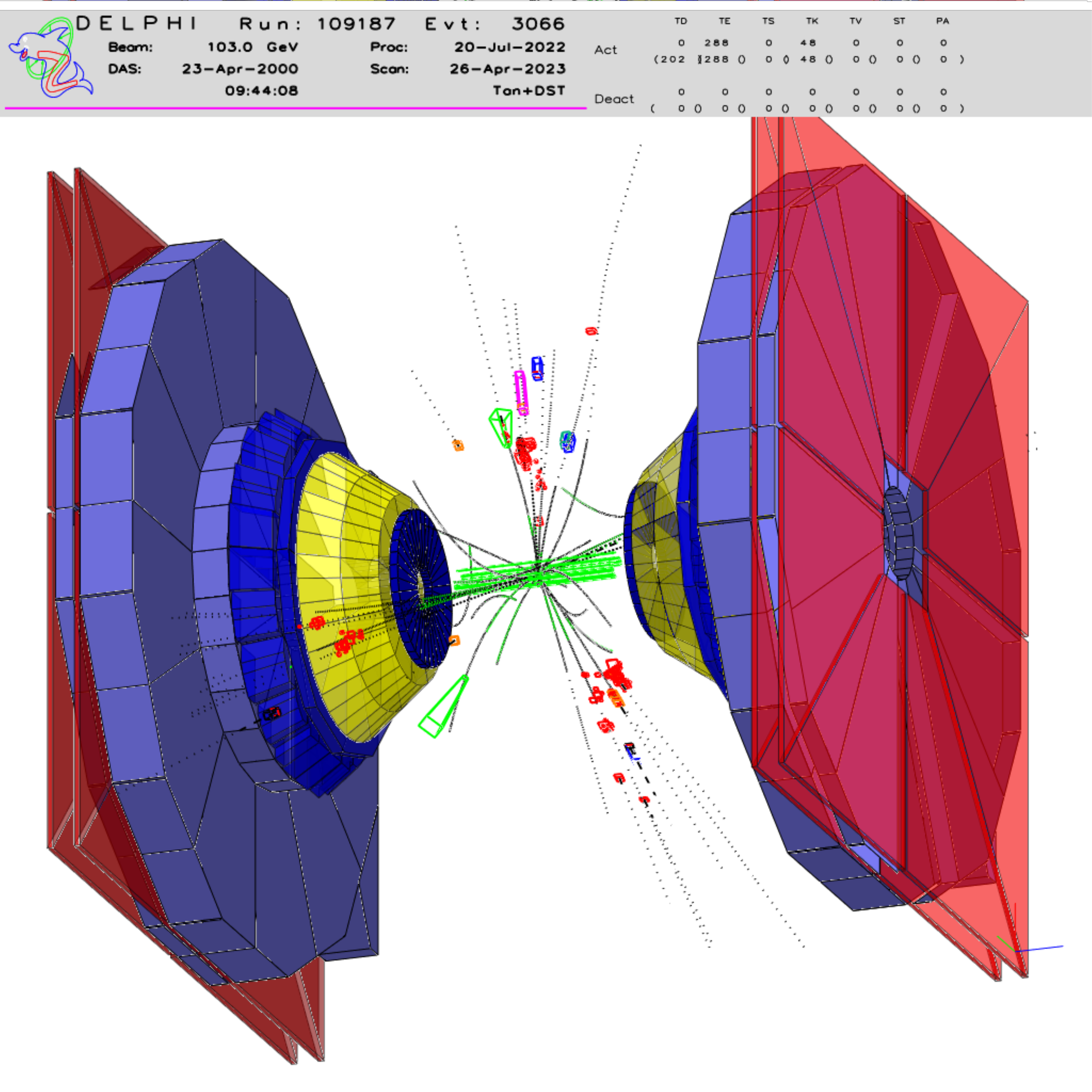,width=0.7\textwidth}
\end{center}
\caption{Example of a DELPHI event, reconstructed from raw data using the revised software stack.}
\label{Fig:delphiev}
\end{figure*}

Figure~\ref{Fig:delphiev} shows an example of a DELPHI event which has been extracted and reconstructed from raw data using the 64bit stack. The validation of the new stack is still ongoing. %and initial results look promising, however, more work is required.
In parallel, DELPHI provides containers and instructions to run old binaries inside these containers. There are also older SLC6 binaries archived on CVMFS which can be used from a CERNVM image which is shared with OPAL.

On the longer term, the OpenData initiative is a possible option, albeit only for educational purposes. Work on extracting data samples and converting them into an appropriate format has started in 2022 and is currently ongoing.

DELPHI documentation, including scanned and OCR processed technical and internal notes dating back to 1982, are available in the CERN Documentation System (CDS).
On analysis preservation, some analysis codes were preserved along with their output n-tuples. These can be useful as additional checks to validate the revised software stack. 

\subsubsection{L3}

L3 legacy data are preserved and stored in EOS at CERN (/eos/experiment/l3/) in several formats. Data in the original L3 compressed format (DVNs) were also converted into \root files and are stored in EOS as well. These \root files can be easily processed using existing \root libraries and utilities. Detailed documentation on the content of these files exists, although is not publicly available yet. The latest L3 policies allowed public use of these data under supervision by L3 members before publication, in order to ensure proper interpretation of the information. 

\subsubsection{OPAL}
\label{LEP:OPAL}

OPAL has preserved its data and complete software stack, with the exception of the event display, and kept it working up to CentOS8. CERNLIB and in particular ZEBRA are needed for access to the data. A 32-bit SLC6 based VM has been defined, but creating new instances does require the option to boot and install SLC6 based nodes, showing the weakness of relying on VMs for data archiving. Keeping the whole software alive and porting it to newer OS versions appears to be a safer approach, but also requires porting the required external libraries (CERNLIB, ZEBRA, etc.). For OS versions newer than SLC6 the availability of the revived CERNLIB, see chapter~\ref{chap:cernlib}, is essential. The data has been migrated to EOS, and the software environment is being migrated from AFS to CVMFS. HBOOK/PAW based n-tuples exist for some specific types of analysis. A proper validation suite is not available.
The documentation has been stored in CDS. 
 
\subsubsection{LEP data and Key4hep}
\label{key4hep_section} 
As seen in the previous sections, the approach for accessing LEP data is very different from one experiment to the other, and is typically limited to the remaining specialists. In some cases, reduced tuples have been extracted and provided to the external users in simple formats; but the process is not automated and it is difficult to imagine a generic solution for FCC-$ee$ based on such an approach.

To address these difficulties, it has been suggested that a possible solution could be connected to the recently started key4hep project~\cite{key4hep}. Key4hep aims to create a common low maintenance, flexible and user-oriented ecosystem of software components, inter-operating through a common event data model, EDM4hep, providing the language for transient and persistent storage.

Key4hep is the framework used for FCC-$ee$ data processing at all steps, from generation to analysis. Converting the LEP data, at least the ones used for analysis, in EDM4hep would have automatically enable the user community to access the LEP data using FCC tools. The migration would have a significant impact also for long term data preservation of LEP data, helping to achieve the FAIR data principles by improving on: Accessibility, detaching from library and OS obsolescence; Interoperability, promoting a single standard framework; Re-usability, requiring less specific expertise. 

A preliminary investigation has shown that the migration could be feasible, at least at level 3, i.e. to ``perform complete analyses when the existing detector reconstruction and simulated data sets are adequate for the pursued goal''. The approach, which could be fully automated, would be to go through XML-like intermediate files, follow by conversion to EDM4hep.

The migration process would require resources and investment, but the return could be huge, not forgetting that the current high interest from EW/Higgs factories studies may provide a unique possibility.

\subsection{H1 and ZEUS}

%Here some introduction on HERA/DESY efforts

HERA is to date the unique electron- and positron--proton collider, operated at energies up to 318~GeV at DESY, Hamburg, from 1992 to 2007. The main collider experiments H1~\cite{H1:1996jzy,H1:1996prr} and ZEUS~\cite{Holm:1993frx} accumulated data samples corresponding to a total of $0.5$~fb$^{-1}$.
Computing needs and data sizes of the HERA experiments are relatively small, compared to other experiments on the DESY IT analysis and storage system host. On best effort, DESY IT will continue hosting the data needed for analysis and offer computing resources for analysis. It has shown that the HERA experiments integrate well with the other HEP experiments in the common NAF cluster. Thus, they only put a small additional support and resource load, and at the same time benefit from the continuously evolving computing infrastructure. The DESY NAF supports container technologies, so that also freezing of experiment code is possible. 

DESY IT benefits from the HERA DPHEP experience in other projects leveraging the use of FAIR data and sustainable analysis of data.

\subsubsection{H1}

The H1 experiment~\cite{H1:1996jzy,H1:1996prr} recorded a unique data set of lepton--proton collisions at HERA in the years $1992$ to $2007$.
The complete RAW collision data comprises around $75$~TB, the set of compressed DST data amounts to about $20$~TB and the analysis level files are about $4$~TB.
A sizeable and universally employed software stack for the processing and analysis of the H1 data also exists, which was initially developed in the years $1988$-$2012$.
This is comprised of a series of core software packages in {\tt Fortran} and the object--oriented analysis core framework, {\texttt{H1oo}\xspace}~\cite{Steder:2011zz}., which is written in {\tt C++} and based on {\tt ROOT}.
Following a DHPEP level 4 preservation policy, the H1 Collaboration continues to maintain these data, all related software, including simulation and reconstruction code, as well as all relevant documentation on the data, MCs, software, detector, operation, meetings and collaboration life. A continuously updated webserver\footnote{https://www-h1.desy.de} provides access to all resources for the collaboration members and for external visitors.
%
%%%
\begin{figure*}[hhh]
\centering
\includegraphics[width = 0.40\textwidth]{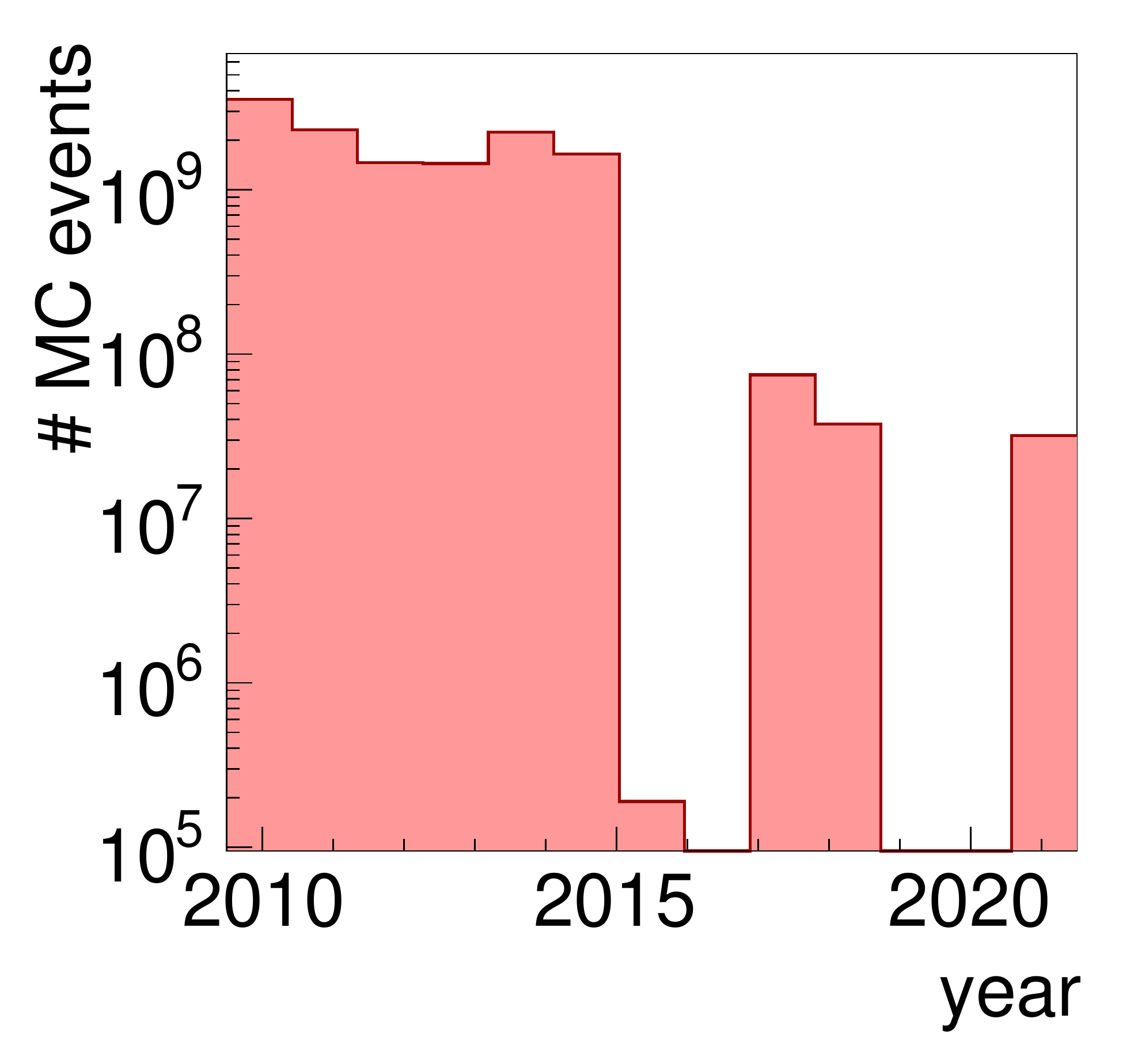}
\hfill
\includegraphics[width = 0.54\textwidth]{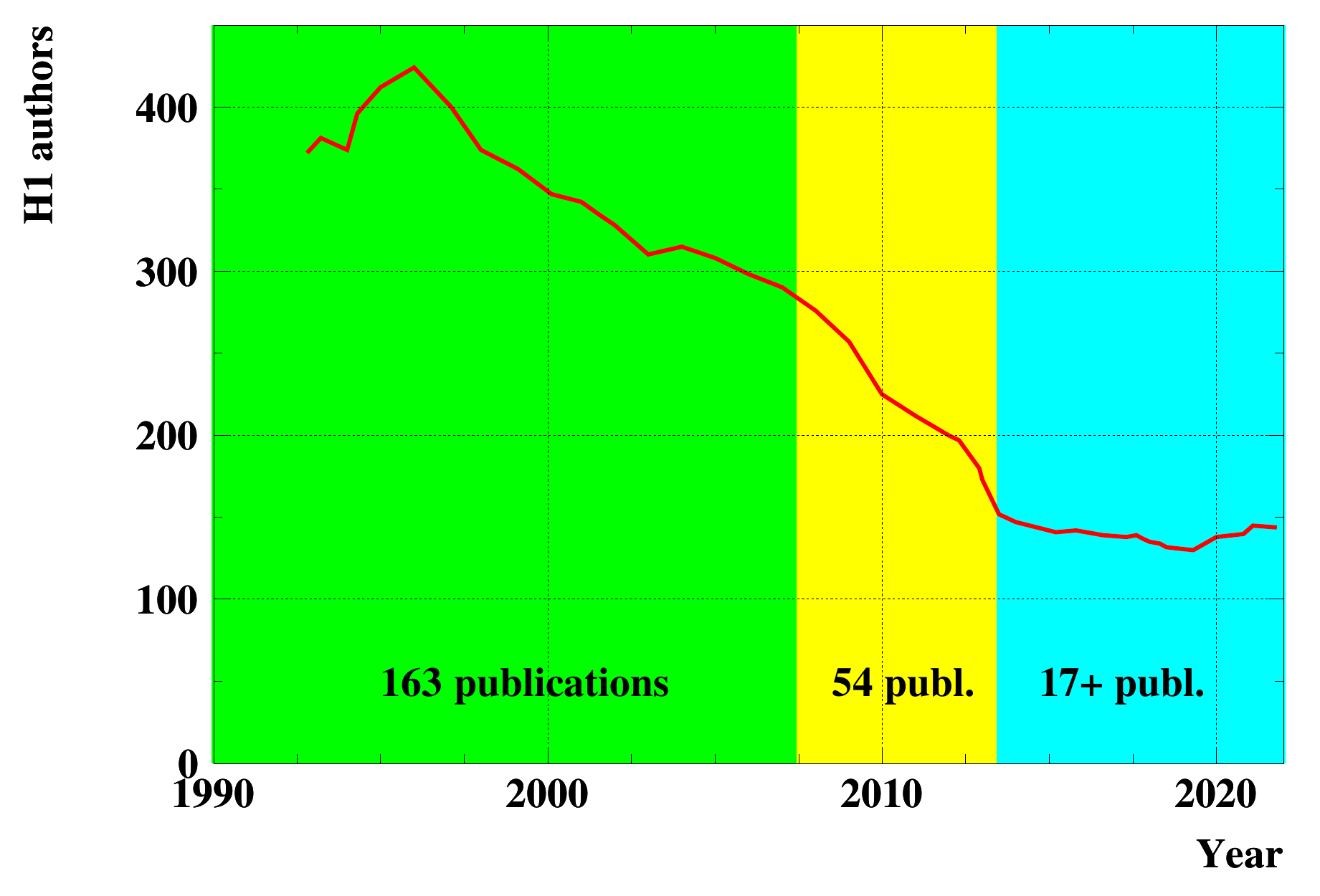}
\caption{
Left: Number of Monte Carlo events produced centrally by the H1 Collaboration.
The years without MC production are related to a change of the computing environment, or no MC requests.
Right: Number of H1 authors 
is increasing since 2019 due to retained analysis capabilities and new interest in $ep$ physics. The colored areas indicate the data taking period (green), the period with active funding (yellow) and the period under the new collaboration agreement in \emph{data preservation mode} (cyan). The number of corresponding publications is also indicated. 
}
\label{fig:H1}
\end{figure*}

The H1 Collaboration works under a renewed collaboration agreement with individuals as the only entity, and it is coordinated by a spokesperson, two deputies and a scientific secretary, and
DESY acts as host laboratory.
Since 2012, an additional twenty H1 papers ($9\%$ of the total) have been published, and presently several data analyses are ongoing or started recently.
In 2021, a new Monte Carlo production campaign with about $4\cdot10^7$ events was performed to support one of the ongoing analyses and the full capability of the present software environment was proven.
%
%When H1 moved into a "preserved operational mode" in $2012$, the majority of dedicated resources %were replaced with centrally managed DESY--IT computing infrastructure, which is shared with %other experiments.
%
Access to the data, software and internal documentation is granted through a DESY computing account, which also provides access to computing resources for data analysis activities.
Data, software or internal documentation are not planned to be made public because of missing or unresolved copyrights of large parts of the software and documentation.
However, the H1 Collaboration is open to new members and saw even an increase in signing authors for the recent publications compared to those from a few years ago.

Any documentation is provided through a dedicated webserver at DESY, where more than 12\,000 digital documents and notes, as well as about 4000 presentations of internal meetings, and the original internal webpages are maintained. 
Relevant documentation in other locations, like institutional web spaces, private repositories, or workgroup servers, were carefully migrated to the common storage.
The offline documentation is stored in 150m of shelve space in the DESY library archive.
%%%

Since 2012, several software migrations were performed including the migration of the entire software and environment to 64--bit operating systems, up to Scientific Linux 6 and CentOS7.
Previous stable software releases are kept in a central repository and can be used for bit-level validations or for production within containerized workflows.
A comprehensive update of the software and analysis frameworks was performed recently~\cite{Britzger:2021xcx}, and is briefly summarized in the following.
%%%

Whilst the full data analysis capability and flexibility was retained, after one decade, the overall status of the H1 software was shown to have a few shortcomings from the present point of view.
The programming languages and standards such as {\tt C++98} were found to be unattractive for young people to learn and liable to slow down new developments and data analysis efforts.
In addition, outdated dependencies such as  \root 5 complicated the usage of modern data analysis techniques and tools.
Furthermore, the risk of incompatibility due to different compiler standards or different interfaces such as MC event generators only increases with time.
As no externally maintained package repository was used, new packages had to be provided manually and the compilation and maintenance still required a profound and detailed knowledge about the specific structure of the H1 software, as well as some insight into the historic development.
The renewed structure of the H1 software stack and environment~\cite{Britzger:2021xcx} allows for an easy transition to even newer platforms, like  AlmaLinux 8 or CentOS8 or 9, and are kept up-to-date with DESY's central Linux platforms.
%%%

Considering the above and further arguments, the status of the H1 software and the data analysis capabilities was revisited in $2020$.
All core software packages have since been successfully migrated to a modern computing platform, based on \texttt{amd64} (x86-64) {\tt CentOS7}, using the {\tt GNU} $9.2$ compiler.
%
%These software modules are required for data access, data processing, reconstruction, simulation, visualisation and, of course, high--level data analysis.
%
Remaining external dependencies were updated to the latest releases and are now obtained from the \texttt{LCG/AA} package repository~\cite{1462660,Roiser_2010}, greatly reducing the number of H1 specific solutions.
The common object--oriented data analysis framework {\texttt{H1oo}\xspace} is now based on {\tt \root 6} and supports the {\tt C++17} standard. 
%
%This framework facilitates the communication within the collaboration members, provides a common standard, and additionally provides highly valuable documentation.
%
%Some example programs and a few selected full high--level analysis codes from previous publications are prepared for newcomers.
%
An online code documentation for {\texttt{H1oo}\xspace} is available, whilst interactive analysis in {\tt C++} through {\tt CLING} is now also possible, as well as for the first time in {\tt Python} (v3).
Several benchmark analyses codes were migrated, like those of jet production~\cite{H1:2014cbm,H1:2016goa} or inclusive DIS cross section measurements~\cite{H1:2012qti}, and serve as a valuable and validated starting point for new data analyses.
%%%
The programs and libraries are now provided to the members of the H1 Collaboration on shared global file systems for convenience.
All H1 software packages are now maintained in {\tt git} and new build instructions for a complete rebuild of the entire software stack have been prepared.
Container solutions have also been implemented for backward compatibility and software tests.
%
%The modernisation of the H1 software is described in more detail in a recent %publication~\cite{Britzger:2021xcx}.
%%%
New Monte Carlo event production campaigns with the full detector simulations and run-dependent conditions can be, and are, performed at the DESY computing cluster(s) and using the modernized H1 software stack. See Fig.~\ref{fig:H1} for a historical summary of the H1 MC production campaigns.

Many H1 data analysis activities are still ongoing to this day, and new analysis projects are beginning due to the uniqueness of this scientific data set.
There is an increasing interest of the HEP community in $ep$ scattering, in particular by physicists from EIC, and this interest is also reflected in the recent addition of new collaboration members. 
This is directly seen from the number of H1 authors, which, after a minimum number of 130 in 2019, is now steadily increasing again, see Fig.~\ref{fig:H1}.

By carrying out modifications to the software architecture, H1 is confident in providing high quality data analysis capability of the unique HERA data in the future, using modern analysis tools, like deep-neural networks for reconstruction or unfolding~\cite{H1:2021wkz,Arratia:2021tsq}, and recent programming languages on state of the art platforms.

\subsubsection{ZEUS}

The ZEUS experiment~\cite{Holm:1993frx} has recorded an equally unique data set of electron-proton collisions, with a 
general purpose detector, partially complementary to H1. All data taken after 1995 up to the end of data taking in 2007 are preserved and continue
to be analyzed. 

The ZEUS data and knowledge preservation project~\cite{Malka:2012kqa} was internally started in 2006 and was generalized through founding contributions to the 2012 DPHEP study group document~\cite{DPHEPStudyGroup:2012dsv} and the subsequent DPHEP collaboration agreement~\cite{DPHEP:2015npg}.

Some of the physics goals to be pursued further were formulated at the workshop on Future Physics with HERA data in 2014~\cite{Bacchetta:2016rdn, Geiser:2015pwp} and many of these have steadily been implemented since. They were and are being complemented by results on new ideas which were not yet in the focus of attention at that time, often implemented by new groups, e.g. from the EIC, Heavy Ion and theory communities, recently joining the collaboration.

An extensive documentation on the ZEUS detector, analysis techniques, available data and MC generated samples is available from the dedicated web sites\footnote{\href{www-zeus.desy.de}{www-zeus.desy.de} and \href{www-zeus.mpp.mpg.de}{www-zeus.mpp.mpg.de}}.

\begin{figure*}[tb]
\centering
\includegraphics[width = 0.49\textwidth]{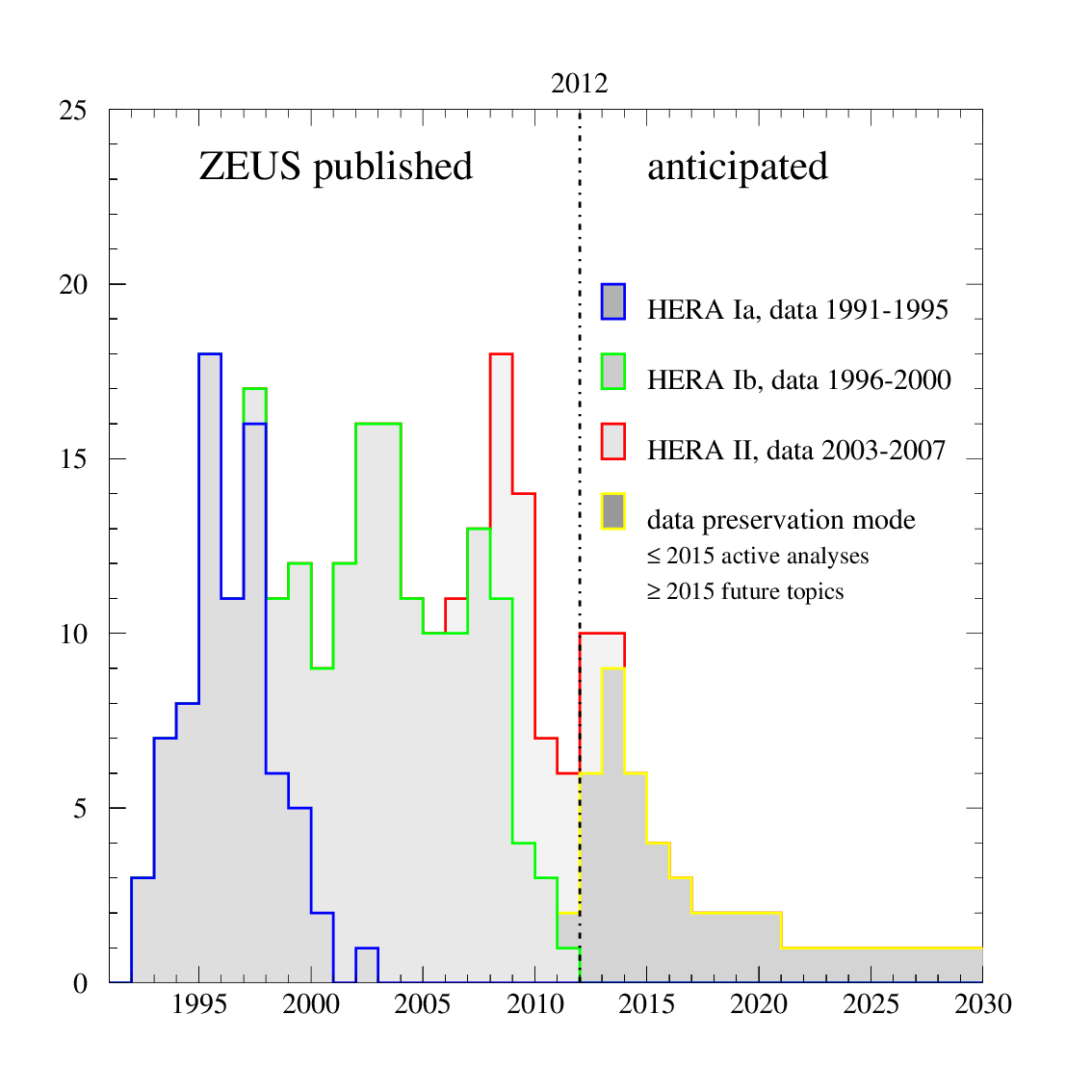}
\includegraphics[width = 0.49\textwidth]{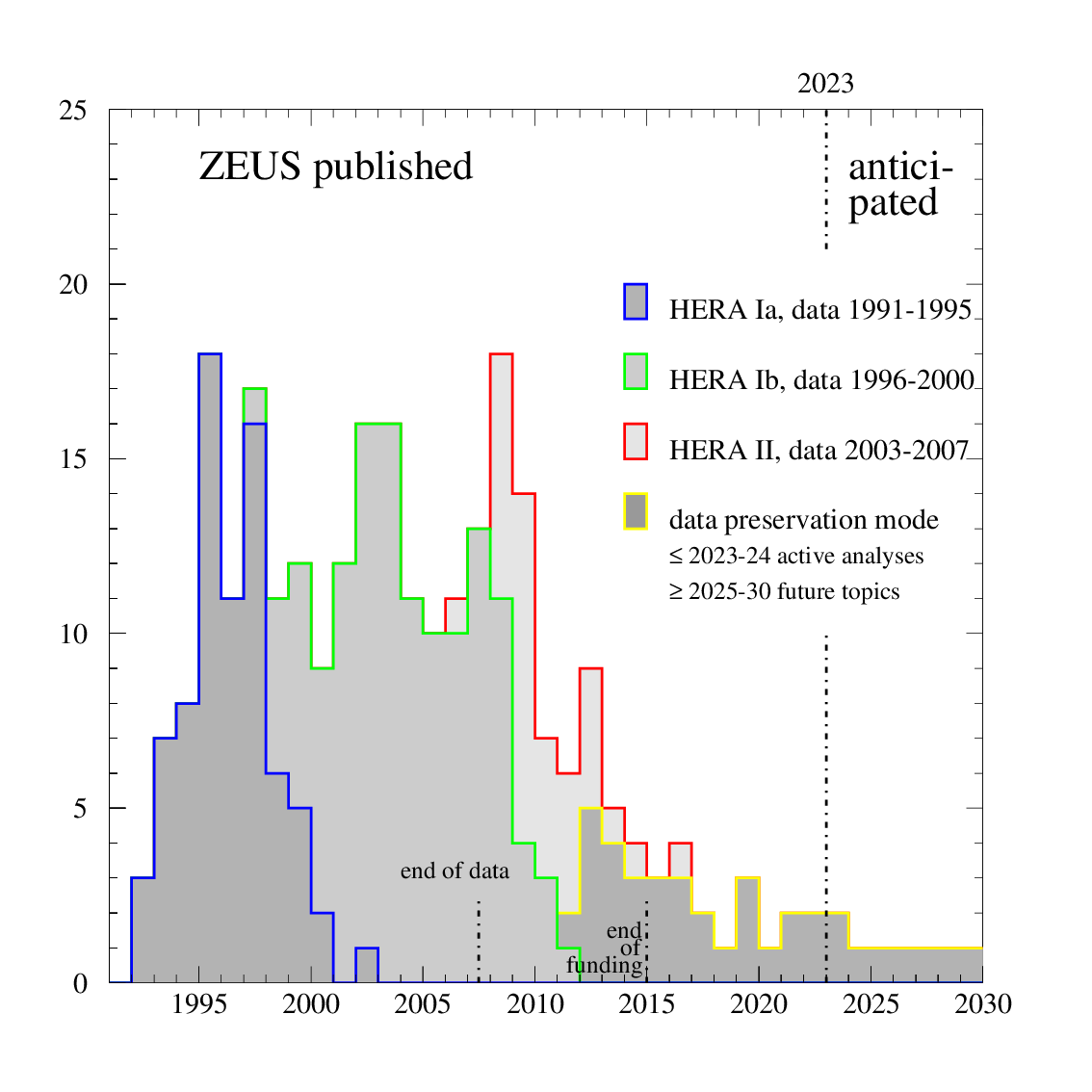}
\caption{Number of ZEUS papers published and anticipated to be published per year. Original 2012 version (left), compared to current 2023 version (right).}
\label{fig:ZEUSpapers}
\end{figure*}

Like almost all HEP papers, the ZEUS physics papers are stored on arXiv~\cite{arxiv} and accessible on INSPIRE~\cite{inspire}. All recent papers were published in Open Access mode. Fig. \ref{fig:ZEUSpapers} shows the evolution of the expectation on the number of publications with time as projected at the time of the study group document~\cite{DPHEPStudyGroup:2012dsv} compared to the latest update of this projection including the publications actually achieved.  The two agree quite reasonably, enhancing confidence into the further projections for the next decade. Integrating up to that time, about 10\% of the total ZEUS physics output (already at $>$5\% now) is expected to have resulted from the dedicated data preservation effort, after the end of funding, without which 80--90\% of these results would never have existed. This is a great return for the $<$1\% additional investment originally made (or rather diverted from the base budget) for data preservation. 
The ZEUS data preservation strategy lies half way between the `level 3' and `level 4' preservation strategies as defined by DPHEP~\cite{DPHEPStudyGroup:2012dsv}. All the data are being preserved at `level 3', i.e. in the form of reconstructed and calibrated low level basic objects (calorimeters deposits, tracks, lepton candidates, ...), as well as higher level composite objects (e.g. jets, missing ET, particular meson decays, ...) which can also be rebuilt from the available basic information. These data comprise so-called `common n-tuples' created from the 360 M real events recorded by the ZEUS detector between 1996 and 2007.

They are stored in a unified flat \root n-tuple format which can be read and processed with almost any past or future version of the \root software package, or any other package providing a \root data format interface such as e.g. root-numpy, pyroot, RDataFrame or uproot. Consequently so far essentially no software maintenance updates were needed since the original setup in 2006, there is no need for virtual machines or containerization, and in this respect the ZEUS analysis software is always as modern as \root or any future backwards compatible successor (note that \root itself is backwards compatible to earlier HBOOK/PAW formats dating all the way back to the late 1970s through a simple `h2root' converter).

All simulated data sets available up to the time of the end of funding in 2014 are also stored in the same format. The relations between the MC generated events and the description of the simulated processes is documented in dedicated internal ZEUS web pages and the lists of files belonging to each generated MC sample are also stored in a standalone {\tt sqlite}  database. The generation of small additional simulated data sets including detector simulation (level 4, through an encapsulated and/or containerized approach) is possible~\cite{Verbytskyi:2016vtw} and has been used successfully. Dedicated codes needed  for the {\tt zevis} event display and for automated handling of the {\tt  cninfo}  data base  (both very useful but not crucial) are available as well in \github~\footnote{https://github.com/}.  

All basic real and simulated data, with a total size of about 250 Tb, are stored and made available in two different geographical locations (DESY/Hamburg and MPP/Garching) using dcache technology~\cite{Krucker:2015nzi}, and can be used through respective generic computing facilities, as detailed above for the DESY case. Optionally, they can also be accessed through the ZEUS Grid instance. 

Currently, the data are accessible only to ZEUS members or individual ZEUS associates, but plans exist to release at least part of the data as Open Data through the recently funded German national PUNCH4NFDI project~\cite{punch4nfdi}.    

In summary, the ZEUS data preservation project, in line with the HERA data preservation project as a whole and the general DPHEP strategy, is successful. Large parts of this program have been implemented as foreseen with very limited dedicated resources; (more resources would have allowed and would still allow the success to be even bigger). More than 30\% of the total HERA physics results were produced in the 14 years since the end of data taking, and more results are coming. This was made possible by the essential support of the host laboratory DESY during the final phase of funding and the continued IT support, and is currently sustained to a large extent by person power originating from external sources.
%\subsubsection{HERMES?}

\subsection{\babar }
The \babar\ detector operated at the PEP-II asymmetric-energy electron-anti-electron storage ring at the SLAC National Accelerator Laboratory, and collected physics data from October 1999 until April 2008. The data were collected mostly at the Y(4S) resonance at 10.56~GeV/c$^2$, the B-anti-B meson production threshold. CP violation and B physics were the main research program but, at that energy, the cross sections for tau-anti-tau lepton and c-anti-c quark production are of comparable magnitude, making of \babar\ a flavor factory, able to access also lepton physics and charm physics. Continuum (u,s, and d quark production) and ISR physics are also accessible. The \babar\ data set also includes a scan of the Y(nS) resonances, in particular \babar\ has the largest Y(3S) data set ever collected and, to date, there are no plans for taking data at the Y(3S) peak at Belle-II (or any other current experiment).

\begin{figure*}[tp]
\centering
\includegraphics[width = 0.75\textwidth]{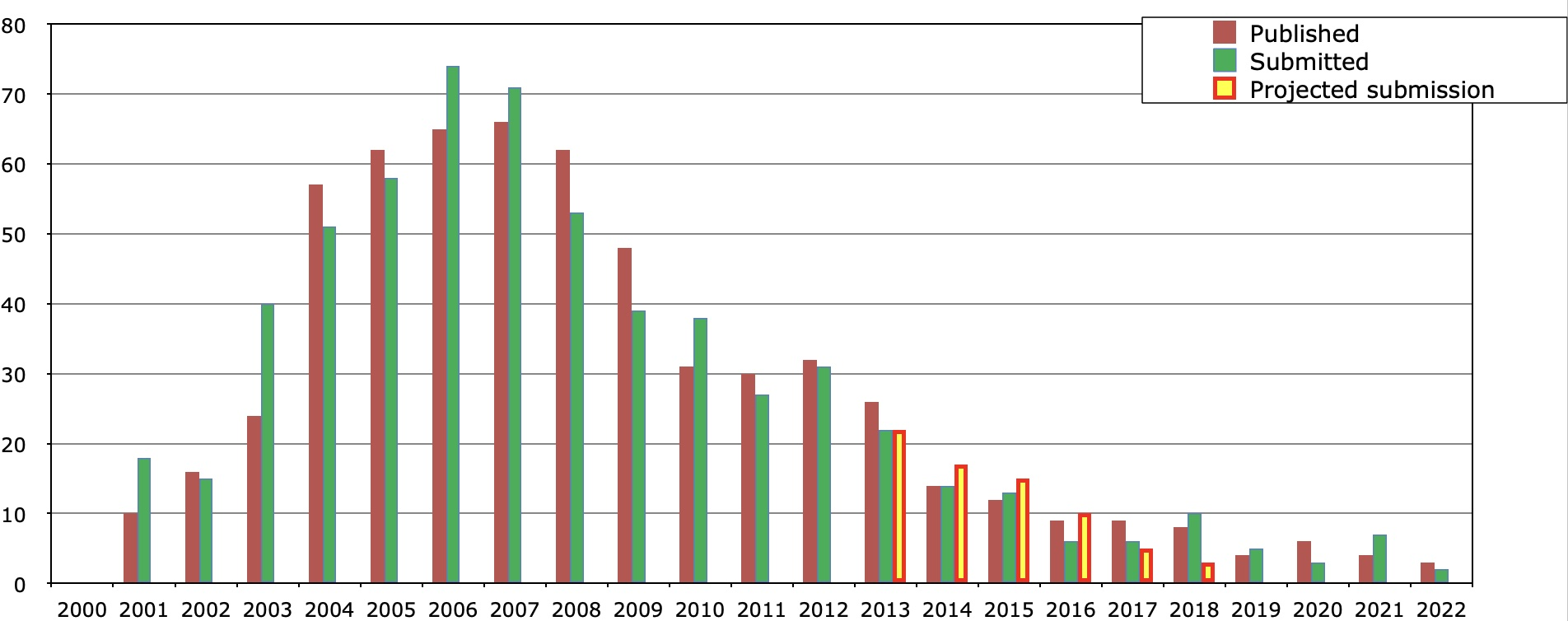}
\caption{\babar\ submitted (green) and published (red) papers per year. In 2012 predictions for submissions (yellow) were made for the years 2013 to 2018. In 2012 it was predicted that no analysis would run after 2018. }
\label{fig:BabarPubl}
\end{figure*}

\babar\ data are in \root format, and the software is written in C++, using an Object Oriented approach. While the software is based on 32-bit, it can run in 32-bit mode also on 64-bit architectures when the 32bit compatibility layer is available. The 32-bit binaries in the latest release work on SL6 64-bit and could potentially be ported to a CentOS7 or CentOS8 derivative, but there is no plan currently to do so due to the limited availability of expert man power. Currently, we operate the software and perform physics analyses using SL6 virtual machines. With a fully working software release and the raw data, \babar\ could aim to a ``Level 4" preservation model, the main problem at this stage being man power and limited resources for the needed infrastructure.

In February 2021, SLAC support for \babar\ stopped and the Collaboration worked on porting the infrastructure on other, more sustainable platforms. In particular:
\begin{enumerate}
    \item Data
    \begin{itemize}
        \item Processed data, real and simulated (1.2PB), hosted at GridKa (Germany), remotely accessed via XRootD for physics analysis
        \item CC-IN2P3 (France) hosts a full copy of the data (2PB including the raw data, for recovery purposes only, MoU in place until 2025 with renewal option)
        \item CERN (Swiss) hosts a full copy of the data as well (transfers still on-going but nearing completion)
    \end{itemize}
    \item Collaboration tools are hosted on a variety of platforms now
    \begin{itemize}
        \item InspireHEP private collections for archiving internal documentation of published results
        \item CERN indico for meetings
        \item CERN e-groups for information exchange and to control access to internal meetings on indico
        \item Caltech hosts general mailing lists to reach all \babar\ members and associates
        \item Google tools for membership and active analysis tracking
        \item \href{http://heprc.phys.uvic.ca/}{The HEP Research Computing (HEP-RC)} group at the University of Victoria, Canada, hosts web based documentation (historical discussion forums, HTML pages, and wiki)
    \end{itemize}
    \item Analysis resources: The University of Victoria HEP-RC group hosts servers for providing user accounts, storage areas, access to the data, and resources for analyses. Servers have been provided by the HEP-RC group and the \babar\ group at the University of Texas at Dallas. In addition, a few machines previously used by \babar\ at SLAC have been moved from SLAC to Victoria to be integrated into the new analysis system.
\end{enumerate}

While the Collaboration retains the ownership of the data, we also support a \href{https://babar.heprc.uvic.ca/www/join_BaBar.html}{\babar\ Associates Open-Access Program} which allows any interested individual or group to easily  join \babar\ and receive the full support of the Collaboration and access data, software, and all internal documents. This approach has already been successful in a number of cases.

The \babar\ Collaboration continues to exploit the rich data set and continues to produce a wealth of world-class results. To date, the \babar\ Collaboration has published 598 papers and 2 more are currently accepted for publication. About 20 new analyses have started in the last 4 years. Figure~\ref{fig:BabarPubl} presents the predicted and current status of \babar\ publications. The figure illustrates the impressive scientific productivity of the collaboration in the long term as well as the reliability of the preservation model and its predictions over one decade. 

\subsection{Belle I and II}
The Belle I experiment~\cite{Belle:2000cnh} collected data set corresponding to about 1~ab$^{-1}$~\cite{Belle:lumi}  at the electron-positron collider KEKB, situated at the KEK laboratory, Tsukuba, Japan. 

The Belle II experiment~\cite{Belle-II:2010dht} at the upgraded SuperKEKB accelerator has started the data taking with the full detector in 2019. Literally, the Belle II is the successor of the Belle I experiment. By the end of 2021, the experiment accumulated around a data set corresponding to an integrated liuminosity of 270 fb$^{-1}$~\cite{Belle-II:lumi}, starting to challenge the Belle I precision. Indeed, Belle II provides already new physics results~\cite{Belle-II:2019qfb, Belle-II:2020jti}.  

However, some of the Belle data such as Y(6S) on-resonance data are world-unique data sets. Therefore, the collaboration decided to preserve all  RAW and mDST (miniDST) data from Belle I. Data Summary Table for physics analysis) data sets for the Belle II experiment and make these Belle data and software accessible to the Belle II collaborators at least until the Belle II data precision exceeds the Belle I precision. 

Currently, these Belle I data are stored at the KEK central computing (KEKCC) system, which is also the main computing system for the Belle II experiment. Although Belle II adopted the distributed computing system, Belle data is at present only accessible from KEKCC. Belle software has been frozen in 2009, except for some patches. The Belle software was originally developed with SL5, but it was migrated to SL6 and CentOS7 because it is necessary for the signal MC production even now. In parallel, the Belle data I/O tool was developed and integrated in the Belle II software framework, so that the Belle II collaborators can read and analyze them only with the Belle II software in which the recent analysis tools are available. 

Thanks to these data and analysis preservation efforts, Belle I physics analysis activities are still ongoing  ten years after the end of the data taking. 
\begin{figure*}[tp]
\centering
\includegraphics[width = 0.95\textwidth]{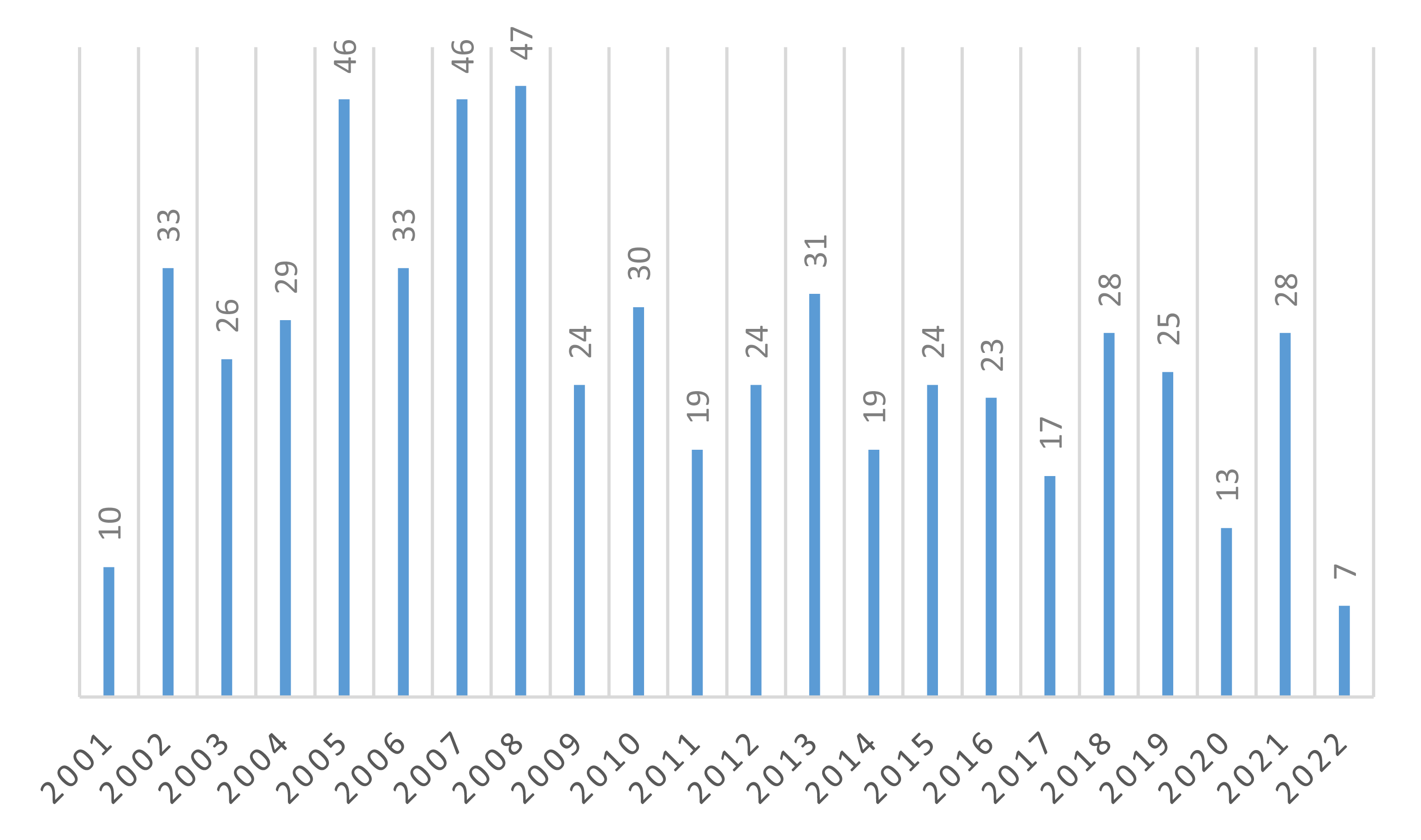}
\caption{Number of the published paper per year from Belle.}
\label{fig:BellePaper}
\end{figure*}

Figure~\ref{fig:BellePaper} shows the number of published physics paper per year from Belle since 2001. The collaboration remains active in publishing new physics results, with a rate of about 20 papers per year.

For the future Belle data preservation, the data need to be migrated  from the current KEKCC to a new system, scheduled to be installed by summer 2024. Data and analysis preservation beyond 2024 is considered, but the detail is not decided yet.

Concerning Belle II experiment, a task force has been launched to evaluate the strategy of the data and analysis preservation in the long term. Under this activity, are discussed the future computing model both in the post-SuperKEKB-running period and post-Belle II experiment lifetime, the period of time for accessibility of the preserved data, the necessary analysis infrastructure, the estimated cost and effort, in line with the DPHEP recommendations.

\subsection{CDF and D0}

The CDF and D0 experiments were multipurpose detectors at the Fermilab Tevatron proton--antiproton collider. They completed data collection in 2011 and jointly participated in the Run II Data Preservation Project (R2DP), described in detail in Ref.~\cite{Amerio:2017kda}. Each experiment has a total preserved data set of approximately 10 PB, and Fermilab has migrated both of them to LTO8 media. Both experiments produced over 500 publications, including two for CDF and three for D0 in the past two years.

The goal of R2DP was DPHEP Level 4 preservation through at least the end of life of Scientific Linux 6 (2020). Containerized environments in conjunction with CVMFS allow for a longer timeline, as long as appropriate SL6 containers are available. CDF code built for SL6 has also been validated to run on SL7. Fermilab officially declared CDF and D0 “completed” experiments in 2021, though the CVMFS repositories and data sets remain available for additional analysis. Fermilab’s Core Computing Division has also begun producing public webpages for “historical” experiments, of which CDF will be an early example.

\subsection{PHENIX}
PHENIX is the largest of the four experiments that have taken data until 2016 at the Relativistic Heavy Ion Collider (RHIC) at Brookhaven National Laboratory (BNL). 
The main challenge in the area of Data and Analysis Preservation (DAP) in PHENIX is that the preservation effort started in earnest fairly recently (2019) and and in the final stages of the lifecycle of the experiment. PHENIX collaboration continues to conduct analyses since the end of the data taking, producing 13 journal articles in 2019-2020 and a substantial number of conference contributions. This work is being done against the background of a gradually diminishing number of active contributors and a large and complex software environment maintained by the facility, as well as a sophisticated analysis apparatus. Over the past two decades the legacy web services became obsolete in terms of both technology and content, and hard to maintain. Knowledge management is, as for other experiments discussed in this report, a substantial challenge.

In order to address these issues, the PHENIX Collaboration has undertaken an effort to put in place Data and Analysis Preservation procedures and practices including
\begin{itemize}
    \item Use of Docker containers to preserve specialized and/or legacy computing environments and enhance software portability. Because of the large volume of the software stack, parts of it were refactored into packages deployed on CVMFS. In addition to fully fledged images of the complete stack (currently kept in a private registry at BNL), lightweight images containing specific versions of \root and other basic software were also created and placed on Docker Hub to simplify access.
    \item Leveraging REANA as a mechanism to capture final stages of select analyses for preservation, validation and user training. Testing of simple serial workflows has been done, and work is underway to implement more complete analyses using more complex workflows which require the use of CWL (Common Workflow Language).
    \item Active supervision and management of the materials created and submitted by the Collaboration to the CERN HEPData portal (Level 1 in standard DAP nomenclature), with a broad team Involvement.
    \item Joining the CERN OpenData portal and using that platform to host self-contained packages which include PHENIX special purpose limited datasets and basic examples of analysis software (Level 2 in standard DAP nomenclature).
    \item A vigorous team effort to migrate PHENIX research materials from legacy information systems approaching end-of-life to a robust and well maintained digital repository, opting to use the Zenodo instance at CERN. There are currently approx. 500 items migrated in this manner.
    \item Development and deployment of a new Collaboration website for easy access to curated materials including those obtained from legacy resources, optimized for long-term stability and ease of maintenance and based on the static website generator technology.
\end{itemize}

Common across all these work areas is the strategy of using community-developed and supported tools, frameworks and services while keeping in-house development to the absolute minimum. In that regard, collaboration with DPHEP provides the most value and is central to the achievements of DAP in PHENIX. In doing this work, PHENIX enjoys the support and contributions from the SDCC computing facility at BNL.

%\subsection{Minerva/Dune (FNAL)}

\subsection{MINERvA}
MINERvA~\cite{MINERvA:2015yej} is a neutrino scattering experiment at Fermilab that recorded data between 2009-2019.  The MINERvA collaboration has published more than thirty neutrino interaction cross section measurements~\cite{MINERvA_WEB} that will inform and tune interaction models for future oscillation experiments such as DUNE and HyperKamiokande.  To ensure that its data is usable into the 2030's, the MINERvA collaboration began a data preservation project in 2019~\cite{Fine:2020snd}.  The project has three components:
\begin{itemize}
    \item Preservation of the high- and low-level reconstructed objects in a ROOT-based analysis n-tuples corresponding to the entire MINERvA dataset.
   \item     A software library known as the MINERvA Analysis Toolkit (MAT) with utilities for transforming the ROOT-based tuples into cross section measurements.  
   \item     A second software library (MAT-MINERvA) based on MAT that can be used to reproduce existing MINERvA analyses and form the basis for new analyses.  
\end{itemize}
As of 2022, the two software libraries have been developed and are available on \github~\cite{MAT,MATMINERvA}, and the datasets are available on Fermilab-hosted DCache servers.

\subsection{BES III}

BESIII~\cite{BESIII:2020nme} at the BEPCII accelerator is a major upgrade of BESII at the BEPC for the studies of hadron physics and $\tau$-charm physics with the highest accuracy achieved until now. The peak luminosity of the double-ring e$^+$e$^-$ collider, BEPCII, is $10^{33}$~cm$^{-2}$s$^{-1}$ at center-of-mass energy 3.78 GeV. 

BESIII is an unique experiment currently operating in tau-charm energy zone in the world. It started to collect data in May 2009, and its end of data acquisition will be extended to 2030. 

The data preservation of BESIII will follow LEVEL 4 of DPHEP. These data is expected to be preserved for another 5-10 years after the end of data acquisition. IHEP computer center have preserved about 4 PB raw data, 1 PB data of other types on tapes which are managed by IBM 3584 library. Migration from CASTOR to EOSCTA (EOS based CERN Tape Archive) has been initiated at September, 2021. Since there is a large gap between CASTOR 1.x and EOSCTA, real data copy is unavoidable in the process of migration. After the migration, the bit preservation technology stack will bealigned with CERN-IT and good practices of CERN can be reused in the future. Tape upgrade from LTO4 to LTO7 format will be finished in this migration at the same time. BESIII Offline Software System (BOSS) is the offline data processing software system of BESIII. It is developed on top of GAUDI. It releases stable versions every year while the latest version is 7.0.8. Currently, OS version of physical computing node has been upgraded to CentOS 7. Earlier BOSS versions and their depenencies are kept in CVMFS and earlier OS version (Scientific Linux 5 and 6) are provided by Singularity containers. Environments of releases preceding  $6.5.5$ can only be recovered in virtual machines. 

There are several HEP experiments at IHEP which have overlap with BESIII in terms of physicist, software developers and IT supports. Their DPHEP policies will generally follow the experience of BESIII. In the near future, their will be 2-3 neutron/photon sources running simultaneously at IHEP and its south branch. DPHEP policies of these new scientific facilities require more attentions. In 2019, the project of ``National high energy physics data center" was proved by Chinese Ministry of Science and Technology. This project will provide support of manpower and funding for the DPHEP at IHEP. With support of this project, IHEP-CC can make further explorations on software technologies of open data, outreach and reusable data analysis.

\subsection{ATLAS, CMS, LHCb, ALICE}

The Large Hadron Collider started the data taking in 2010 with two multi-purpose (ATLAS and CMS) and two specialized (LHCb and ALICE) experiments. The collaborations have adopted since several years data preservation policies both for the ongoing runs and also as specifications for the future upgrades. A clear synergy, pointed out from the very first DPHEP reports, between preservation and open science is successfully practiced by all LHC experiments, as described in this section. Furthermore, transverse projects focused on generic solutions for data re-use and re-analysis are focused on the LHC, as will be described in the next chapter. The data sets are most sensibly grouped by data taking periods, commonly called "Runs" (such as the ongoing Run 3 from 2023 to 2026), are particularly favourable to structuring the data towards open access and preservation. 
%(here some introduction on LHC efforts towards DP and open data)

\subsubsection{ATLAS}

The ATLAS approach to data preservation is informed by a larger policy framework designed to ensure the long-term impact of the collaboration’s physics program during and beyond its lifetime. Thus, while the raw data generated by the experiment are preserved, ATLAS also invests in analysis, metadata, and software preservation, resulting in a spectrum of data and software products tailored towards a number of audiences. All public data products released by ATLAS aim to adhere to FAIR principles by leveraging community web infrastructure to aid in locating relevant data (findable) and using standard access protocols (accessible). When possible, community-developed common data formats are chosen (interoperable). Finally, data and software products are explicitly designed to be input for new research rather than an archival preservation of prior investigations (reusable).

Data generated by the ATLAS experiment are categorized according to the levels defined in a previous DPHEP report~\cite{DPHEPStudyGroup:2012dsv}, to which varying preservation and access policies are attached. Level 1 data represents data released publicly alongside a published result by the collaboration. First and foremost this includes the publications themselves, which are made available as Open Access under the SCOAP3 initiative. In addition to the paper itself, additional data are released publicly on community cyberinfrastructure such as HEPData~\cite{HEPdata} to facilitate reuse by physicists. This includes digitized tables and figures from the publication, such as yield tables and theory parameter bounds derived from the data. Additionally, data are released  pertaining to the analysis design that allow the approximate reimplementation of the data analysis procedure, such as tabulated selection efficiencies, key multivariate observable definitions such as neural networks and decision trees, or cutflow data. More than half of ATLAS analyses currently include HEPData records. ATLAS also routinely releases collaboration-validated software implementations approximating the actual event selection as Rivet routines~\cite{RIVETanr} and SimpleAnalysis classes~\cite{SimpleAna}. The current Rivet analysis coverage of ATLAS is 35\%~\cite{RIVETfo} (the highest of any tracked experiment). For SimpleAnalysis, ATLAS provides the SimpleAnalysis framework which is released as Open Source software~\cite{SimpleAnaSoft}, and more than 35 search analyses have been included already. As a key data product for re-use, ATLAS also releases the statistical models underlying the search or measurement (see e.g. Refs.~\cite{ATLAS:2021kak, ATLAS:2019oik}). This captures all systematic effects and allows external researchers to reproduce the statistical procedure at full fidelity or  re-use ATLAS results in follow-up studies such as global statistical inference. The focus of public Level 1 products is the use of long-term, stable human- and machine-readable formats. The Level 1 records are assigned stable identifiers (DOIs) and can be individually cited by third-party researchers.

Augmenting the public Level 1 data, ATLAS preserves detailed internal publication-specific information through two complementary approaches. A comprehensive metadata record in its internal GLANCE~\cite{Maidantchik:2008zza} database captures the full analysis life cycle from inception to final journal publication and collects diverse metadata such as analysis team membership, presentation, approval reviews, code repositories, physics keywords, etc. In addition, a growing number of analyses are preserved as a software product for future reuse and reproducibility. Here, analyses are described using a declarative computational workflow language, using parameterized task templates and Linux Container Images. The latter are able to capture not only the analysis software but its full set of dependencies (OS, language runtimes, compilers, etc). A particular use-case is the reinterpretation of analyses in terms of alternative signal models using RECAST~\cite{Cranmer:2010hk, ATLAS:2020viz}. Here a tight integration with the REANA platform developed by CERN IT~\cite{REANA} was developed.

ATLAS does not host Level 1 data directly but relies on the continued development and maintenance of the archival cyberinfrastructure, most crucially HEPData, to provide this service to the community.

At Level 2, ATLAS compiles special-purpose data sets targeted for outreach and education beyond the HEP community and releases them publicly on the CERN Open Data Portal (see e.g. Refs.~\cite{ATLAS:2020oda, ATLAS:2020odb}). The data sets are deliberately simplified in both content and format to facilitate broad usage in non-academic settings, minimizing technical boundaries. As a consequence these data are not suitable for research usage. Alongside the data sets, a range of software examples such as interactive notebooks, data analysis macros, or visualization code are prepared and maintained (see e.g. Ref.~\cite{ATLAS:2020odc}).

Reconstructed event-wise data sets beyond the scope of a single analysis are categorized as DPHEP Level 3. These represent the main data used by ATLAS members wishing to prepare a physics result and are available internally in a range of data formats. The data are produced from raw data sets using the ATLAS reconstruction software Athena, which is made available as Open Source and whose development is publicly accessible~\cite{ATLAS:soft}. In addition, Athena is citable through a corresponding Zenodo record~\cite{ATLAS:zenodo}. A given release of the software is packaged and preserved independently from the source repository using industry standard formats for global distribution. In some cases, container images of the software are available~\cite{ATLAS:soft1}. Both observed collision events and simulated events are processed through tightly controlled versions and configurations of the software, and for any given data set, detailed metadata and provenance information is stored in the ATLAS Metadata Interface (AMI). Based on these metadata, reproduction of derived data sets is straightforward if necessary. To ensure a coherent data set of all data produced during a data-taking period, the reconstruction software only undergoes performance improvements, while major changes and algorithmic improvements are incorporated during large-scale reprocessing campaigns. Such reprocessing of old data with new software is a key step towards internal data preservation and allows ATLAS to maintain consistent multi-run data sets that may be processed by a single software release. This was recently undertaken to unify the format of the Run 2 data set with the upcoming Run 3 data set.

For public access to Level-3 data, ATLAS has implemented the guidelines of the broader CERN Open Data Policy~\cite{ATLAS:2020odd}. In this scheme, ATLAS will follow a staged collaboration-defined data release schedule that begins to publicly release Level-3 collision and simulation data within five years after the close of a data-taking period and ensures a full release of the complete data set after the close of the experiment. The time delay allows for an initial exploitation of the data by the collaboration and ensures that the released data have benefited from all algorithmic improvements and calibration studies of the collaboration. The data are planned to be released in a calibrated format designed also to be used internally for the bulk of ATLAS analyses. Alongside the data, ATLAS releases software to perform analysis at the same level of fidelity, including the assessment of systematic uncertainties. The full ATLAS data analysis releases are Open Source~\cite{ATLAS:soft} and additionally are preserved regularly using Linux Containers~\cite{ATLAS:soft1}.

In addition to the bulk release of collision and simulation data in a fully calibrated format, ATLAS has also produced several special purpose data sets for R\&D purposes. Two examples are the data sets prepared for the TrackML project~\cite{Elsing}, a public data challenge for the development of machine-learning based tracking algorithms, and the Higgs ML challenge~\cite{ATLAS:2020ode}. Similarly, a data set was recently released for the training of machine-learning-based calorimeter simulation~\cite{ATLAS:zenodo1, ATLAS:2020odf}. Such data sets may consist of fewer or only a subset of events but include more low-level information than the bulk release.

Level 3 data are released through the common CERN Open Data Portal with CERN as the host laboratory assuming custodial responsibility over the released data. Similarly, the corresponding software and containers are hosted on CERN infrastructure. It is anticipated that any public metadata associated with each analysis and data set would likewise be hosted on common infrastructure, for example in the CERN Analysis Preservation Portal (CAP)~\cite{CAP}. Like HEPData, these infrastructure components are crucial to the success of the Open Science program of ATLAS but not directly maintained by the collaboration.

The DPHEP Level 4 category describes the raw data from the detector before reconstruction. These data are not immediately usable for physics research and are rarely accessed by collaboration researchers. They are therefore not released publicly during the lifetime of the experiment. These data are, however, preserved and serve as input for future reprocessing by the experiment, and they will become public after the close of the experiment for archival and historical purposes.

In preparation for the LHC Run 3, the ATLAS software underwent a major evolution, from multi-process to multi-threaded code. This allowed significantly improved memory usage in the ATLAS software~\cite{ATLAS:2021sft}. To ensure efficient use of the ATLAS Run 2 data, it was decided that prior to the Run 3 data taking, the Run 2 data set and corresponding Monte Carlo samples should be reprocessed. This reprocessing campaign was completed in 2022; however, it required a major effort, in particular for the data and the changing conditions throughout the data taking period. Periodically reprocessing the data in this manner has proven key to avoiding a loss of interoperability in internal data sets or additional unnecessary analysis complexity.

For the ATLAS collaboration, the preservation of its data and the search for perennial means for its efficient scientific use is of great importance. The intrinsic complexity of the data, metadata and data structure and the software to analyze it pose a major challenge to finding practical data and analysis preservation solutions. The collaboration believes that the current level of effort invested in data and analysis preservation and the planned effort to support the release of open Level 3 data are appropriate to match the current scientific needs. Extending the scope of the ATLAS data preservation efforts would require additional human and computing resources from external sources.

\subsubsection{CMS}

The CMS experiment is addressing the data preservation and reusability through regular open data releases, putting into action the experiment’s Data preservation, reuse and open access policy~\cite{CMS:2020od}. The policy defines the collaboration’s approach for maintaining data collected by the experiment usable in long term. It was first approved in 2012, as the first of such policies in the HEP domain, and it has been updated in 2018 and 2020. CMS has mandated a Data Preservation and Open Access Group responsible for managing the implementation of the policy.

The CMS data policy is in agreement with the CERN open data policy, and like all LHC experiments, CMS publishes results in Open Access journals. Additional data, at so-called Level 1, to facilitate immediate re-use and the combination of these results are provided through, and archived in the long-term, by trusted third parties such as HEPData. At Level 2, simplified data formats for several levels of immediate re-use such as limited analyses, education, and outreach are available, accessible, and preserved through the CERN Open Data portal~\cite{COD}.

The Level 3 consists of data that can be used to reproduce published analyses and perform new analyses, and CMS was the first HEP experiment to release data of research quality and has done pioneering work in the domain since 2014. The CMS data releases take place regularly through the CERN Open Data portal. As defined in the CMS policy, CMS will normally make 50\% of its data available 6 years after they have been taken. The proportion will rise to 100\% within 10 years, or when the main analysis work on these data in CMS has ended. However, the amount of open data will be limited to 20\% of data with a similar centre-of-mass energy and collision type while such data are still planned to be taken.

The Level 4 consists of raw data, as stored directly after data-taking without further processing. Custodial copies of these data are stored at CERN and at the corresponding custodial computing Tier 1 for each data set. CMS can release small samples of raw data potentially useful for studies in the machine learning domain and beyond together with level 3 formats. If storage space will be available, raw data can be made public after the end of all data taking and analysis

The CMS open data releases contain full reprocessing of collision data from each data-taking period and the simulated data corresponding to these data. They are made available in the format and with the same data quality requirements that analyses of the CMS collaboration start from. The public data are accompanied by a compatible version of the CMSSW software and additional information necessary to perform a research-level physics analysis. Example code and some specific guide pages are provided to explain and instruct the use of this associated information.

\begin{figure*}[tp]
\centering
\includegraphics[width = 0.98\textwidth]{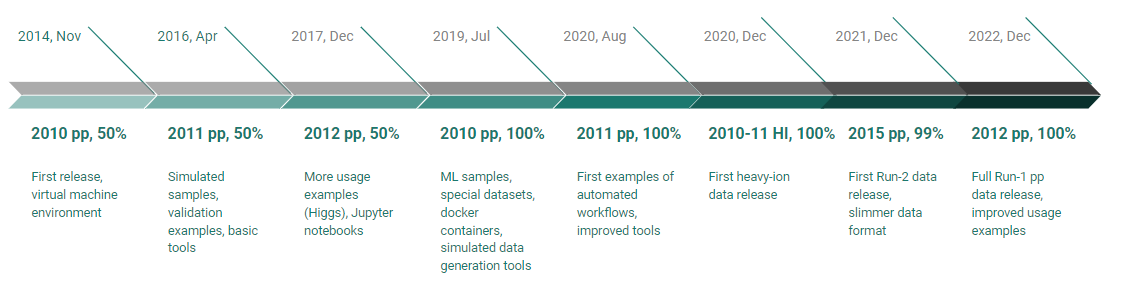}
\caption{CMS data release timeline.}
\label{fig:CMStime}
\end{figure*}

The first CMS data release took place in 2014, followed by regular releases since then, as shown in the timeline in Fig~\ref{fig:CMStime}. At the time of writing of this document, all Run-1 proton-proton data from 2010-2012 and first heavy-ion data are in the public domain, and the first batch of Run-2 data from 2015 has been released. Run-1 data (2010-2012) are in Analysis Object Data (AOD) format and some early special data in RECO format from which AOD is a subset. Starting from Run-2, a slimmer MiniAOD format is in use, and a reduced NanoAOD format averaging to about 1-2 kB per event will also be made available.

The released data are accompanied with rich metadata, describing their quantity and quality, and importantly, the full provenance information. The provenance information records the exact parameters and software conditions used at the time of data-taking, for collision data, and in the event generation and simulation step for Monte Carlo data. The input parameters are recorded also for the subsequent processing steps. This information together with the open-source CMSSW software makes it possible to trace back all data handling, although doing so is certainly a tedious task. 

A virtualised computing environment, compatible with the software version with which the original data can be analysed, is provided and maintained. The CMS provides docker software containers~\cite{CMS:cont1, CMS:cont2}, regularly tested and updated, and Virtual Machine (VM) images, based on the CernVM software appliance~\cite{CMS:vm}. They are part of the data release, as well as any additional information necessary to perform a research-level physics analysis. The additional data products are needed in different steps of the analysis, for example for data selection, as correction factors to be applied to physics objects, or for evaluating the final measurable results in terms of cross sections.

CMS data released through the CERN open data portal satisfy FAIR principles for Findable, Accessible, Interoperable, and Re-usable data and metadata to a large extent. But due to the complexity of experimental particle physics data, the FAIR principles alone do not guarantee the re-usability of these data, and additional effort is needed to pass on the knowledge needed to use and interpret them correctly.

This knowledge includes, first of all, learning the computing environment and software for the first step of data selection. Due to the experiment-specific data format, the first step will almost inevitably be done using the CMS software in a computing environment compatible with the open data. Open data users can download a software container image and run it on their computer, independently of its operating system. Recent developments for Windows Subsystem Linux (WSL2) have also made this feasible in Windows, in addition to Linux and macOS. The CMS open data group has invested a good amount of work in setting up these containers so that the first user experience with the CMS open data remains smooth.

After having set up the computing environment, open data users will need to learn the intricacies of experimental particle physics data. How to select the data of interest, how to identify the particles properly, how to understand the efficiencies and uncertainties in the analysis process, how to estimate the backgrounds, and how to address many other challenges with experimental data. To address these questions, the CMS open data group is putting together a comprehensive CMS open data guide~\cite{CMS:2020oda}. In addition, regular workshops are conducted for physicists interested in research use of these unique public data. These workshops aim to cover the skills needed to get started with the experimental particle physics data from the CMS experiment and get participants initiated with hands-on exercises. Sessions practising running CMS data analysis jobs in a cloud computing environment are also included.

To ease the task of getting properly started with open data analysis, CMS is setting up all open data examples as workflows that can be run an automated way. This also makes it possible to regularly test and verify that all data and computing assets are working as expected. Similarly, work is ongoing to set up test workflows to monitor that reprocessing raw data and regenerating new Monte Carlo data remains possible in the legacy environment.

To encourage wider utilisation of the knowledge embedded in physics analysis work within the collaboration, analysis procedures, workflows, and code will be preserved internally in version-controlled code repositories and software image registries (e.g. CERN \gitlab\footnote{\url{https://about.gitlab.com/}}) which are currently being set up. They will be made available and searchable to collaborators through CERN analysis preservation services~\cite{CAP}. The open data releases can include selected analysis workflows, at an example level. The reproducibility of the preserved workflows can be tested in systems like REANA~\cite{REANA}.

CMS open data are widely in use. The number of scientific publications by open data users external to the CMS collaboration is comparable to that of an internal working group in the collaboration, thus broadening the scientific value of these data. The usage can be monitored through citations to the data records that all have a digital object identifier. However, the counting mechanism for repositories such as Inspirehep is still imprecise. CMS open data also offer the possibility of benchmarking HEP tools in a realistic context, an opportunity that many development teams have benefited from. In addition, they have been used at university-level education, for example, particle physics courses or as a new type of physics laboratory exercise. Simplified data sets derived from CMS open data have been used to enrich the curriculum at the high-school level and in many particle physics outreach initiatives.

\subsubsection{LHCb}
The LHCb collaboration is committed to the community efforts to make LHC data available to the public. Its collaboration board has ratified the CERN Open Data policy ~\cite{CERN-OPEN-2020-013} in 2020. Data releases are made available through the CERN Open Data portal~\cite{COD} and trusted repositories, such as  HEPData~\cite{HEPdata} or RIVET~\cite{RIVET}, and comprise data on three levels of complexity. Publications are open access and routinely accompanied with supporting materials and releases of machine readable results to HEPData, where appropriate.  All plots and tables of LHCb papers are accessible in machine readable form through the LHCb published results pages\footnote{\url{http://cern.ch/lhcbproject/Publications/LHCbProjectPublic/Summary_all.html}}. 

For educational purposes and outreach selected data sets with various degrees of information depth are published on the open data portal. These datasets are usually intended for a specific application, such as the International Masterclasses, and contain heavily filtered information necessary for the respective purpose. 

In December 2021 LHCb has approved the first release of reconstructed data from Run I  of the LHC. For LHCb the corresponding level or complexity has been defined as the output of the stripping or, where applicable, the turbo stream, which is the same level of abstraction available to the researchers inside the LHCb collaboration. The stripped data contains offline selections of several thousand selection algorithms called stripping lines. The stripping lines are grouped into streams, corresponding to similar physics signatures. In the current scheme, releases are done on a stream by stream basis. The Electroweak, Radiative and Leptonic streams have been approved for release so far, corresponding to about 200 TB of data. The release to the open data portal and the accompanying documentation is currently being finalized. 

The software to analyse these data is available as open source \cite{DaVinci} and documented through the LHCb starterkit. However, only limited support can be provided by LHCb to potential external users. 

The development of the open data curation tools, documentation and metadata schemes is organised within the LHCb Data Processing and Analysis (DPA) project. All related efforts invested by collaboration members count as service work. 

To improve the sustainability of these efforts two issues with the currently planned open data release have to be addressed. First, the data volumes scheduled for release by the LHCb collaboration are quite significant and would amount to higher storage requests on the open data portal than those of ATLAS and CMS combined. Second, despite the best documentation efforts, the access to the data provided by current tools, while appropriately flexible for experts, lacks an intuitive interface, that would facilitate the exploration of the data by external users. 

In order to address both issues a new tool, called the n-tuple-Wizard, is expected to enter beta-testing within the LHCb collaboration in May 2022. The basic idea of this tool is to allow external users to submit queries to the LHCb data, which will be processed by the LHCb analysis productions system accessing the original replicas of the data on the grid. For security reasons the users cannot directly submit analysis jobs to the production system but instead are guided through configuring an analysis in a fully interactive web application. This application also provides search features to locate specific processes of interest in the data, which can be visualised graphically as so called decay trees. The application allows to configure a large fraction of the analysis tools available in LHCb to extract quantities of interest (e.g. kinematic variables or particle identification classifier outputs) from the data. The wizard finally generates a job configuration, which is passed to the LHCb analysis production team without further user interference, where it can be validated and finally run on LHCb compute resources. The result is a plain \root n-tuple, which is made available to the user.

When the beta phase is successful the n-tuple-Wizard is planned to be incorporated into the CERN open data portal. Aside from the technical issues involved in this integration, there are a few open questions related to the management of the generated data, which need to be worked out. The Wizard and the analysis production system provide provenance tracking, but schemes to store this information as meta data accompanying the n-tuples still need to be decided.

The concept aims at allowing to make the existing replicas accessible, thus reducing the need for dedicated open data copies of the data. Moreover, the web application provides guidance and documentation on the meaning of the extracted information. In the future standardised (semantic) meta data, describing the content of the generated n-tuples could be generated as well. Finally, the wizard allows fine grained control over which data is made available, beyond the current by-stream release scheme.

% Analysis preservation
LHCb requires a minimal analysis preservation policy for all published analyses. The analysis software is archived on CERN \gitlab in the physics working group workspaces. Input data, which is typically filtered to the n-tuple-level is archived on EOS (Easy Open Storage file system at CERN) or grid storage. Upstream samples and processing steps, such as the offline stripping selections are managed centrally with a dedicated bookkeeping system based on DIRAC.

With the start of Run 3 LHCb has adopted the real time analysis (RTA) strategy based on a complete online event reconstruction. The default data processing path will not include an offline reconstruction step anymore and the output of the high level trigger (HLLT) will be roughly equivalent to the level of abstraction provided by the stripping offline selection in Runs 1 and 2. The concept has already been successfully piloted by the turbo stream data produced since 2015. From the analysis preservation point of view the move to RTA means that large parts of the data preparation are by construction done centrally managed with all necessary configurations becoming part of the LHCb software stack. 

In order to capture the user analysis parts, which typically deal with interpretation of the data, rather than data preparation,  current efforts focus on the Snakemake workflow engine, which has become quite popular within LHCb and is now also supported by REANA. A future vision would be to enable the majority of LHCb analyses to preserve at least the workflow producing the central value of the analysis for deployment to REANA. 

For the capture of meta-information generated during the internal discussions and review a new database system, called Analysis Lifecycle Management (ALCM) is under construction in LHCb, which will eventually replace the current public results pages and offer the possibility to export public extracts of the information to systems like CAP.

\subsubsection{ALICE}

Recently, ALICE together with the other CERN’s Large Hadron Collider (LHC) collaborations has adopted a new policy~\cite{CERN-OPEN-2020-013} supporting a consistent approach towards the openness and preservation of experimental data. Essentially, all LHC experiments share a common strategy for the release of Published Results (Level 1), Outreach and Education Data (Level 2), and Reconstructed Data (Level 3), while Raw Data (Level 4) are not considered usable in a meaningful way outside the collaborations. 

In compliance with the CERN Open Access Policy, all ALICE publications are available with Open Access and data points together with additional information including the analysis code are made public at the time of publication using the HEPData portal~\cite{HEPdata} and Rivet toolkit~\cite{RIVET}. In addition, dedicated subsets of ALICE data for the purposes of education and outreach have been released in the TTree \root data format and are accessible through the CERN Open Data Portal (CODP)~\cite{COD} together with the needed software to analyze them via virtual machine, container technology, and web-applications. The early implementation of the ALICE open access policy for Reconstructed Data led to public availability of 5\% of 2010 Pb–Pb and 7\% of 2010 proton-proton (pp) collisions data sets in Event Summary Data (ESD) format on the CODP. The software availability has been achieved by using virtualization technology based on microCernVM bootable machine image with CernVM-FS client and the related documentation has been published on CODP. 

To best align the implementation of open data with the new software developments, the ALICE collaboration plans to adopt a new data format to make the ALICE Reconstructed Data public. A large conversion campaign of data from Runs 1 and 2 has been performed and the data sets with the new format is available from April 2022. Such a new format, based on the ALICE’s O2 Project, has been developed for Run 3 and it is more compact and more performant with respect to the previous formats. In addition, having all old data sets converted into the new format will ensure the long-term usability of the old data sets with the new Run 3 analysis framework.

Documentation and software availability constitute the other key elements of the ALICE data preservation strategy allowing the future collaborators, the wider scientific community, and the public to analyze data for educational purposes and for eventual reassessment of the published results. ALICE has already put in place procedures including the use of Docker containers and the software deployment on CVMFS for the software continuous integration and the release validations. Such procedures are also the bases of the Data and Analysis Preservation for the members of the ALICE collaboration, while the wider scientific community and the public cannot benefit from these procedures having no access to CVMFS. For a wider range of users, lightweight images containing specific versions of the new Run 3 ALICE analysis framework and other basic software will be also created and placed on CODP to simplify access.

Further developments to achieve the preservation of the different stages of the analyses have been based on CERN Analysis Preservation (CAP)~\cite{CAP}. Tests of simple serial workflows have been performed creating proper JSON files from analysis trains and uploading them to CAP. The option to rerun ALICE analyses in REANA~\cite{REANA} using inputs from CAP and running code within Docker containers has been tested but it requires more detailed studies with the use of Common Workflow Languages (CWL) to describe more complex workflows.

\begin{comment}

\subsection{Future experiments/plans?}

\subsubsection{FCC}

Despite long-term data preservation not being yet a major preoccupation for FCC, the approach adopted to base the software of the project on Key4hep, the common software ecosystem introduced in Sect.[lep], will facilitate the preservation of the FCC data if Key4hep and the related event data model EDM4hep become adopted standards among HEP experiments. Schema evolution and backward compatibility are indeed among the required features of the common data model. This is currently ensured by using \root file format as backend, which has demonstrated keeping files readable through version changes for a span of 25 years.

\subsubsection{EIC/sPHENIX}

\end{comment}

%===========
%\newpage
\section{DP Technologies and projects}
\label{DPsystems}
Although HEP has a long tradition of common software development, in particular for data access, simulation and analysis (e.g.HIGZ~\cite{Bock:1988pd},  PAW suite~\cite{Brun:1988pg, Brun:2011zzb}, \root~\cite{Antcheva:2011zz}, GEANT3~\cite{Brun:1987ma}), the computing systems and the data structures are largely experiment specific. This variety is of course related to an efficient experiment running and originates mainly from the objective of obtaining promptly physics results, within the resources allocated to each experiment. This objective is not always compatible with a common approach that may involve extra resources not freely available during the data taking.

However, this approach induces a fragility towards a long term data preservation, since some systems are "custom made" and difficult to maintain in the longer term.  It may also raise further difficulties when the data is open to a larger audience, beyond the experimental collaboration, because adjustments are needed towards lighter/standard interfaces - otherwise the learning curve for new users may be overwhelming. Moreover, the data preservation ("cold") systems may require new technologies and new methodologies, beyond ("hot") systems used during the experiment lifetime. 

Therefore, as identified also in the early DPHEP documents, there is a clear need for transverse projects, to be exploited by several or even all experiments, potentially proposing new standards and aligning the goals for long term preservation with those of the open and FAIR data. A few examples of such transverse projects are presented in this section. They fall into two broad categories: a) Transverse projects aiming at providing generic solutions for data re-use: HEPData, CERN Open Data Portal, CERN Analysis Preservation, REANA Reproducible Analyses and Archiver, and b) Common tools and methods for DP : Bit migration, CERNVM and CERNLIB.

\subsection{HEPData}

HEPData is the primary open-access repository for publication-related (Level 1) data from particle physics experiments, with a long history going back to the 1970s.  The HEPData project underwent a complete transformation in 2017 to a new platform (\href{https://www.hepdata.net}{hepdata.net}) hosted on CERN computing infrastructure~\cite{Maguire:2017ypu}.  Another transition was made in 2020 to deploy the web application via a Docker image on a Kubernetes cluster shared with the INSPIRE-HEP project.  Funding is provided by the UK Science and Technology Facilities Council (STFC) to Durham University (UK) for staff to maintain the operation of the \href{https://www.hepdata.net}{hepdata.net} site, provide user support, and develop the open-source software (\href{https://github.com/HEPData}{@HEPData} on \github) underlying the web application.  In the past, data preparation in a standard format and upload to the repository were also handled by HEPData staff at Durham University, but now these tasks are delegated to the experimental collaborations.  

Data submitted to HEPData (as YAML) is primarily in a tabular form that can be interactively plotted in the web application and automatically converted to other common formats (CSV, JSON, ROOT, YODA).  The interactive nature of HEPData means that data tables must be kept sufficiently small ($\sim$MB or less) that they can be quickly loaded in a web browser.  In practice, tables with more than $\sim$10,000 rows (for example, a covariance matrix for a measurement with $\sim$100 bins) cannot be practically rendered in a web browser.  However, moderately large tables or non-tabular data files can be attached to a HEPData record as additional resources (in any format), where the original files can be downloaded but the interactive nature is lost.  To avoid the multiple problems caused by the attempted upload of large files, an overall size limit of 50 MB is currently imposed on the uploaded archive file.  Publication-related (Level 1) high-energy physics data that is not suitable for HEPData, due to either being too large or predominantly in a non-tabular format, should be submitted to another data repository like Zenodo, which currently plugs a gap to host HEP data (and software) that does not fit into other repositories.

\subsection{CERN Open Data Portal}

The CERN Open Data portal~\cite{COD} was launched in 2014. The portal manages and disseminates more than two petabytes of open data from particle physics. The majority of the data content comes from the LHC experiments. The data are being released in periodic batches in close collaboration with LHC experiments and their data preservation teams following the published open data policies. This usually implies a certain embargo period that allows for the exploitation of data within the collaboration as well as for the data curation and verification procedures.

The released open data are described as bibliographic records following a custom metadata schema. The portal focuses on describing the usual metadata (title, authors, keywords, year), the technical metadata (file formats, file sizes, file locations and checksums) as well as high-level metadata necessary to understand the data context (how was the data selected, how it can be used, the data semantics). The bibliographic records are minted with a Digital Object Identifier (DOI) to ease data citation and referencing.

\begin{figure*}[tp]
\centering
\includegraphics[width = 0.9\textwidth]{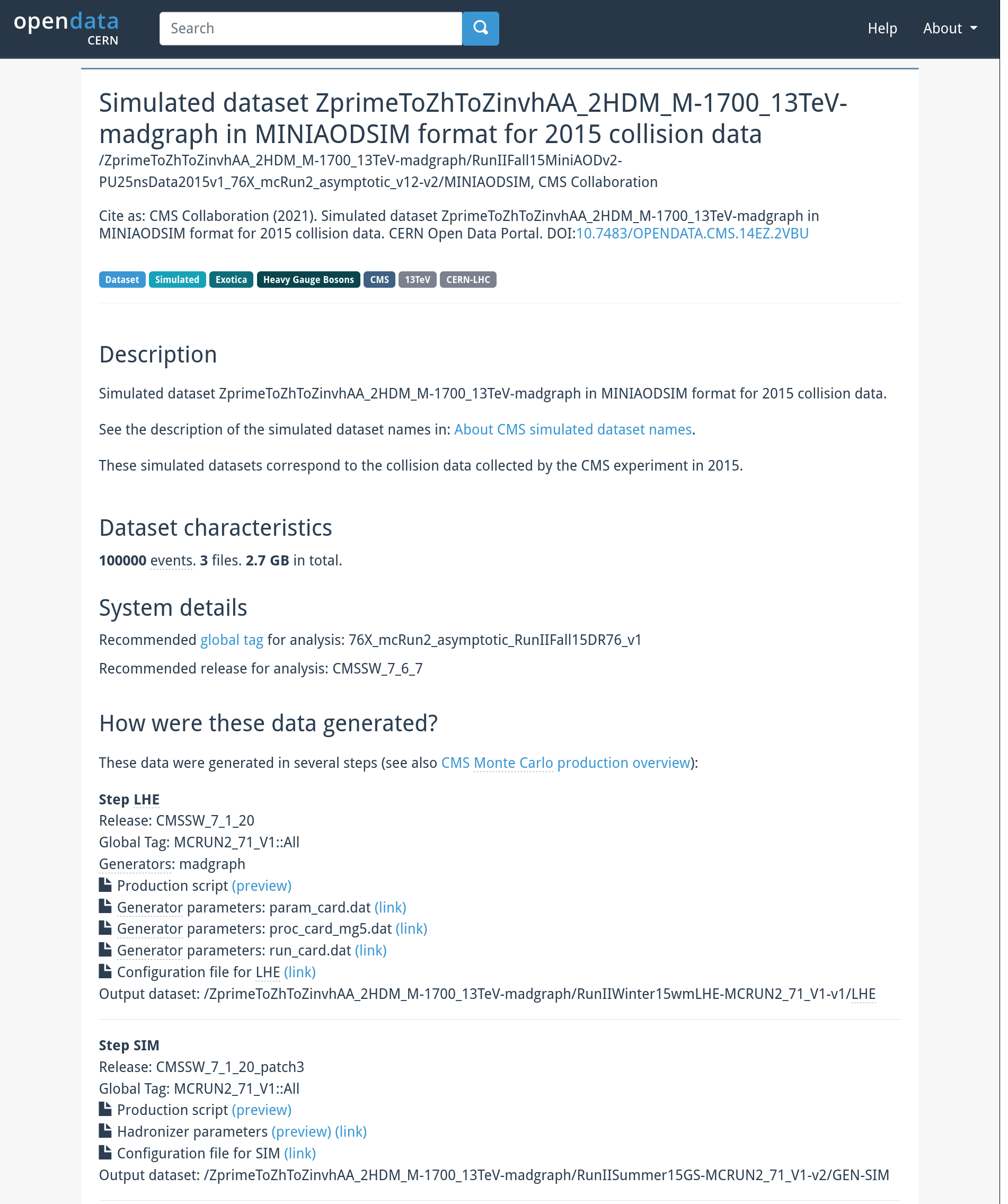}    % lower resolution
\caption{Example of a CMS simulated data set released on the CERN Open Data portal. The full data set provenance information is captured and made available to users.}
\label{fig:CMSod}
\end{figure*}

The CERN Open Data portal content currently contains over 15 thousands bibliographic records describing over 1.2 million files of over 2.8 petabytes. The content represents raw data samples (Level 4),the collision and simulated data sets (Level 3), as well as simplified derived data sets and event display files (Level 2). Whenever possible, the data is accompanied with associated documentation, the data provenance information~\cite{Simko:2020jyw} as well as the corresponding data production configuration files (see Fig.~\ref{fig:CMSod}). The portal also describes software tools and provides data analysis examples together with the associated Virtual Machines and Docker container environments where the data analysis examples can be run. This aims to simplify data reuse for theoretical physicists or machine learning specialists that may undertake data analyses without being necessarily well acquainted with the detailed internal knowledge of the LHC experiment that produced the data.

The curated open data content is being used for both educational and research purposes. The CERN Open Data portal offers embedded event display interfaces allowing particle physics students to visualise detector events and perform basic histogramming on site. The research-grade data is available for download by HTTP and XRootD protocols. A command-line client was developed to allow researchers to automatise download procedures and to verify the integrity of downloaded data for large collision and simulated data sets.

Besides the four LHC experimental collaborations (ALICE, ATLAS, CMS, LHCb), the data sets of relevance to the Machine Learning communities are classified under a separate experiment-independent Data Science collection the interest in which is constantly growing. Beyond LHC particle physics, the CERN Open Data portal nowadays disseminates the OPERA neutrino physics data sets~\cite{DeLellis:2020qch} as well as the first batch of open data from the PHENIX collaboration, demonstrating the widening of the open data coverage from pure CERN LHC particle physics content towards the wider HEP open data content at large.

\subsection{CERN Analysis Preservation}

The CERN Analysis Preservation (CAP) project intends to supply a service at CERN to meet the specific analysis preservation needs of research teams at the Large Hadron Collider (LHC) experiments. Generated at the world’s largest research instrument, through multinational research collaborations consisting of thousands of scientists and engineers, the research data of the LHC experiments is unique, precious and of unparalleled complexity. While the dynamic organizational setting of CERN makes it particularly challenging for the data, software and associated knowledge around a physics analysis to be preserved in a comprehensive reusable manner, the LHC experiments aim to adopt a consistent approach towards the openness and preservation of experimental data. 

CAP was developed to preserve information related to a scientific analysis as it is produced, making it easier to describe, find, exchange and hand-over information in fast paced research environments with highly fluctuating personnel. The service responds to two parallel demands. First, the internal needs of the community, i.e. the LHC experiments, where the high throughput of analyses results in significant challenges in terms of capturing and preserving data analyses and enabling reproducibility of research results, which might be needed at a future point. This is evidenced by the large number of cases where knowledge around analyses was lost once researchers moved on, as some analysis materials had not been properly nor timely made accessible. Second, the external demands: the service allows experimental teams to address the increasingly pressing requirements of funding agencies worldwide who have put in place data management policies regarding research data and knowledge preservation for future reuse and reproducibility.

Developed in close partnership with a diversity of research teams at the LHC, CAP has been designed to flexibly adapt to varying experimental workflows, e. g. building on centralized tools that are already in place or taking over the role of information aggregator. The service facilitates the capture of information about scientific analyses to facilitate reuse and reproduction even many years after its initial publication, permitting to extend the impact of preserved analyses through future revalidation and recasting services. To facilitate the adoption of analysis preservation as a standard part of the scientific process, CAP has been designed to seamlessly integrate into collaboration workflows, so that information can be captured at the earliest stage, and throughout an analysis life-cycle. Designed as a secure environment, experimental teams retain full control of their information and data. CAP enables users to apply the appropriate access restrictions so that users are always in control of when and if their work is shared or published.

Concrete use cases are:
\begin{itemize}
    \item Community analysis information hub: Bring together information from various tools and databases. Search, compare, retrieve and share information.
    \item Validation and reuse: Instantiate preserved analyses and computational workflows on compute clouds to allow their validation or execution with new sets of parameters to test new hypotheses.
    \item Streamlined handover and onboarding: A person having done an analysis is leaving the collaboration and has to hand over the know-how to other collaboration members. A newcomer would like to join a group and requires information on past analyses.
    \item Optimised preservation-publication workflow: Prepare more complex outputs for public releases. Easy information exchange between CAP and public-facing publishing platforms.
    \item Support for Open Science policies: Aggregate and preserve information to comply with any internal or external policy requirements.
    \item Multipurpose reviewing tool: Enable an internal review committee to check and repeat an analysis. % with the information, co
\end{itemize}

Built over several years of focused development and tested across research teams at the LHC, CAP is a mature and robust solution for the preservation of analyses to support data preservation and management to meet funder expectations and research integrity concerns in a world moving increasingly towards open science. CAP has been endorsed by the four main LHC experiments and through its demonstrated success in the highly complex domain of high energy physics, at one of the world’s largest and complex research organizations, the CAP service shows potential for broader application across research disciplines. Furthermore, the technology and tools used to develop the service are not only open source, but are discipline agnostic. More specifically, CAP is built on top of the open source framework Invenio, which already has a wide variety of applications in many disciplines, and all the underlying source code for CAP is openly available on \github.

\subsection{REANA Reproducible Analyses}

REANA is a reproducible analysis platform~\cite{REANA} that aims to facilitate reusable science by allowing researchers to structure and run parametrised computational data analysis workflows. REANA was launched in 2017 as a sister project to ensure the reuse of content preserved in the CERN Open Data portal and the CERN Analysis Preservation framework services~\cite{Chen:2018drk}.

The data analysis process can be generally described by means of specifying input data, the analysis code, and the computational recipe used to arrive at the final results trough a sequence of computational steps. This process can be expressed by means of Directed Acyclic Graph (DAG) where graph vertices represent units of computation with their inputs and outputs and graph edges describe the interconnection of various computational steps~\cite{Simko:2018zzz}. REANA supports several such DAG standards (CWL, Snakemake, Yadage), parses the workflow specification described by the user and dispatches its computational steps to various supported compute backends (Kubernetes, HTCondor, Slurm). The reproducibility of computations is assisted by means of using software containers (Docker, Singularity) that fully encapsulate the original computational environments of each analysis step.

\begin{figure*}[tp]
\centering
\includegraphics[width = 0.95\textwidth]{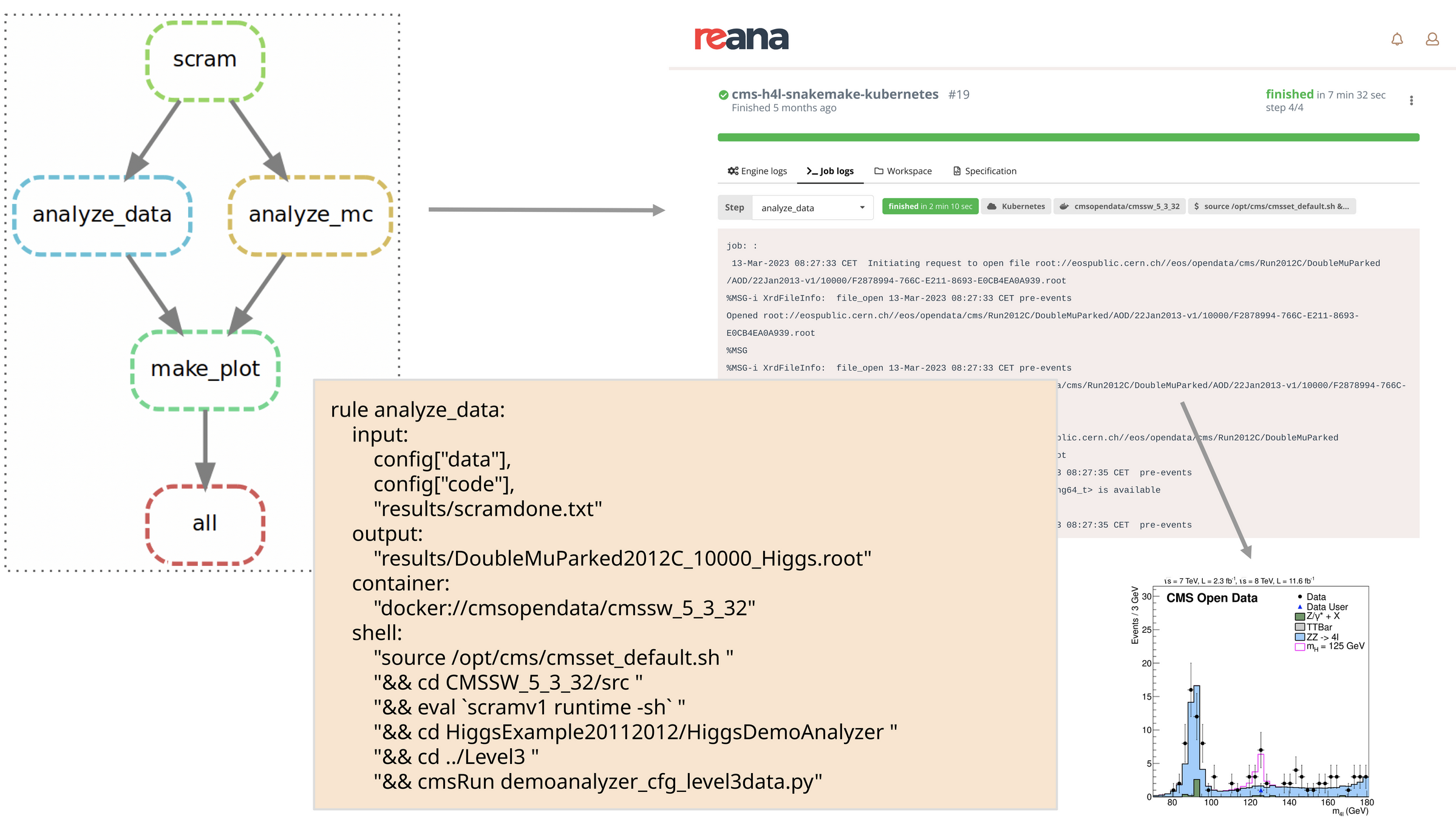}  %reasonable resolution (otherwise overleaf timeout)
\caption{Example of Higgs-to-four-leptons analysis of CMS open data running on the REANA reproducible analysis platform. The analysis consists of four steps and is expressed in the Snakemake workflow specification language. The specification of the collision data analysis step is illustrated in the picture.}
\label{fig:CMSreana}
\end{figure*}

The REANA approach was successfully tested in several data production and data analysis scenarios. For example, REANA was used for ATLAS reinterpretation searches for new physics or CMS reconstruction and jet energy corrections~\cite{Simko:2021gqg}. In the CERN Open Data portal, REANA is used both on the ``data production'' side to ensure the correctness of preserved data sets' provenance information~\cite{Simko:2020jyw} as well as on the ``data analysis'' side by running several data analysis examples such as CMS open data Higgs-to-four-lepton example analysis (see Fig.~\ref{fig:CMSreana}).

REANA promotes early adoption of computational reproducibility principles by researchers. The integration with source code management platforms such as \gitlab~\footnote{\url{https://about.gitlab.com/}} allows researchers to develop and run either full data analysis tasks on REANA or at least test the correctness of analysis workflow after each code change. If an analysis is developed in this ``continuous integration'' manner~\cite{Simko:2021gqg}, the preservation of knowledge associated with the data analysis as well as the future deposit of analysis assets into digital repositories are largely facilitated. REANA therefore complements the data preservation repositories by promoting active ``preproducibility'' of data analyses during the active analysis phase rather than only relying on passive data deposition and subsequent ``reproducibility'' once the analysis is completed~\cite{Stark:2018}.

REANA was developed with particle physics use cases in mind, but the platform profits from synergies with general reproducible data analysis patterns in other scientific disciplines, such as astronomy, bioinformatics and life sciences~\cite{Simko:2018dcl}.

\subsection{ARCHIVER}

The project ARCHIVER~\cite{ARCHIVERWEB, ARCHIVER} develops innovative services for Long Term Digital Preservation of scientific datasets using the Pre-Commercial Procurement instrument funded by the European Commission. R\&D was performed competitively by commercial suppliers, across different implementation phases. The services were developed by Arkivum and Libova co-designed with input from research clusters members: CERN, PIC/IFAE and DESY are members of the ESCAPE cluster~\footnote{\url{https://projectescape.eu/}}, DESY is also a partner in the ExPaNDs\footnote{ExPaNDS is the European Open Science Cloud (EOSC) Photon and Neutron Data Service \url{https://expands.eu/}.} cluster, while EMBL-EBI is a member of EOSC-Life. These research performing organizations deployed use cases from Astrophysics, High-Energy Physics, Life Sciences and Photon-Neutron Sciences. The use-cases driving the ARCHIVER consortium’s need for research and development of innovative data preservation services extended the preservation ecosystems of research organizations to create more dynamic solutions to be deployed primarily by using a hybrid model, on-premise or exclusively in cloud environments, operated by European SMEs that were enhanced and assessed against best practices such of CoreTrustSeal\footnote{A data repository certification \url{https://www.coretrustseal.org/}} and DPC-RAM\footnote{The Data Preservation Coalition (DPC) Rapid Assessment Model (RAM) is a digital preservation maturity modelling tool that has been designed to enable DPC RAM homerapid benchmarking of an organization’s digital preservation capability \url{https://www.dpconline.org/}. }, so that datasets remain FAIR (Findable, Accessible, Interoperable, Reusable) for decade timescales or more.

The services resulting from the ARCHIVER R\&D are addressing the gaps that put data at long-term risk as they prevent the construction and operation of sustainable Trusted Digital Repository services, affecting organizations both large and small who are tasked with being custodians of valuable research artifacts.
Thanks to its outstanding results, the ARCHIVER project has been recognized by the international community and been awarded by the International Council on Archives, category for Collaboration and Cooperation, at the DPC awards ceremony in Glasgow, on the 12th of September 2022.
The model developed in ARCHIVER is considered very beneficial for the research community and contributes to address concerns about sharing and re-use of FAIR data to reproduce research as a pillar for Open Science and long-term preservation of the European Open Science Cloud (EOSC)~\footnote{\url{https://eosc-portal.eu/}.} federated data, by delivering sustainable, production quality long-term data preservation services for user communities that fills a gap in the existing EOSC portfolio.

\subsection{Bit Preservation at CERN}

%{\bf CERN}

During 2021 CERN migrated all archival storage from the earlier CASTOR system to the new CERN Tape Archive (CTA) which consequently now assumes CERN's bit preservation responsibilities. All preservation use cases, including of course the LEP data, were moved into CTA. 

In terms of bit preservation CTA provides essentially the same guarantees as CASTOR. Indeed, the migration was purely a matter of moving metadata and the data on tapes remained unaffected. File access permissions are managed slightly differently in CTA and this triggered discussion on the appropriate model for authorising access to preservation data. A model where a privileged curator is able to maintain a list of reader accounts (using CERN's e-groups system) is being elaborated.

During 2021 CERN deployed a new tape library in a building separate from the computer centre. This library will be commissioned for use with CTA, which will allow dual-replica data to benefit from geographical separation between the replicas.

CTA now holds the data that CERN has received from \babar\ for the purposes of data preservation. The integrity of this archive has yet to be validated by \babar\ and thus the transfer is not yet considered complete.

\subsection{CERNVM}

CERNVM provides a portable platform to develop and execute HEP experiment applications~\cite{cernvm:2011}.
The CERNVM platform consists of a virtual appliance with a minimal operating system (OS) and a distributed file system (CVMFS) that provides a global shared area for scientific applications and conditions data~\cite{cvmfs:2011}.
Both the appliance and the distributed file system distribute data processing environments to the heterogeneous, distributed computing infrastructure used by running experiments.
The CERNVM technology is designed to take software preservation aspects into account and to allow for an easy transition from an actively maintained experiment software stack to the preservation phase~\cite{cernltdp:2016}.

The CERNVM appliance provides both an image for a full virtual machine (VM) for use with hardware virtualization technology such as KVM and Amazon EC2 as well as a container image for use with container tools such as Docker and Kubernetes.
Full VMs give access to hardware emulation technology, so that software stacks compiled for deprecated CPU architectures can run on contemporary hardware, albeit at a significant performance cost (often $\times$10 and more slower).
Container virtualization has a negligible performance overhead; it relies, however, on a stable interface from the Linux kernel to user-land applications.

Retrofitting the CERNVM technology to software stacks from the LEP experiments, the approach has shown to bridge more that 20 years.
A CERNVM environment with a CMS software stack from 2011 to process LHC run 1 data was created with minimal effort, since the CVMFS and CERNVM infrastructure has been in use from the start by LHC experiments.

\begin{comment}

\begin{figure}[tb]
\centering
% TODO \includegraphics[width = 0.7\textwidth]{figures/CERNVMaleph.png}
\caption{ALEPH event display in CERNVM.}
\label{fig:CERNVMlep}
\end{figure}

\end{comment}

Key in the CERNVM approach is the split between the minimal appliance (the OS platform) and the experiment software directory tree provided by the CVMFS network file system.
The na\"{i}ve use of virtualization technology, in contrast, which bundles OS and application software in a single VM or container image, is easily susceptible to container management problems.
Due to the complexity and variety of experiment application software, the na\"{i}ve approach results either in huge images (e.g., 100GB+ for all ATLAS releases) that are difficult to distribute or in a proliferation of special purpose containers that are difficult to book-keep.
Software stacks provided through CVMFS, on the other hand, are well-maintained by the experiments' software librarians and the necessary binaries for any given task are loaded on-demand at runtime.
As CVMFS is a versioning file system, software stacks installed on CVMFS are automatically preserved remain available for future use.

Several lessons were learned from exercising CERNVM and CVMFS for the preservation of LEP and LHC experiment applications:
\begin{enumerate}
    \item It is important to distinguish between \emph{scientific} applications (e.g., event generators, detector simulation, histogramming tools) and \emph{system} applications (e.g., web clients, data access software). 
    In contrast to scientific applications, system applications generally do not freeze well. 
    This is due to the fact that system applications often interact with the outside world or at least the (virtualized) hardware.
    For instance, web clients may unsuccessfully try to connect with an outdated SSL protocol to HTTPS servers.
    Hard disk utilities may be subject to integer overflows on today's volume sizes.
    
    \item As a result, it is important to identify network dependencies of data processing workflows before freezing.
    There may be dependencies, for instance, to online databases (such as conditions databases), grid configuration, or data access servers.
    A stable frozen software environment should only use the POSIX file system as a means to interact with the outside world.
    The experiment software as well as all input and output data should be stored on network file systems, which can be mapped into virtualized environments.
    In CERNVM, CVMFS is used to store software and detector conditions data and EOS is used for experiment data.
    
    \item Lifting authorization barriers (such as X.509 certificate authorization) has been useful in the context of data preservation efforts.
    Authorization protocols are particularly ill-suited for freezing due to sudden upgrades of protocols or software following the revelation of vulnerabilities.
    Also, there are often expiry mechanisms in security protocols such as certificate lifetimes.
    Software and data on CVMFS is publicly readable by design; providing the data on public EOS servers instead of protected shares significantly simplifies the provisioning of historic data processing environments.
\end{enumerate}

\begin{figure}[tb]
\centering
\includegraphics[width = 0.45\textwidth]{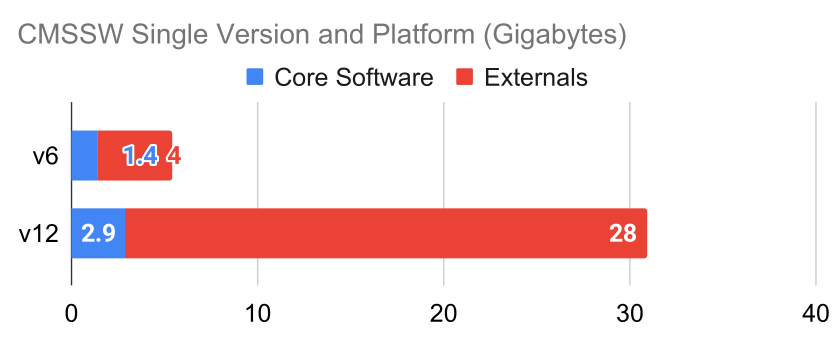}
\caption{Growth of the CMS software stack in approximately 5 years.}
\label{fig:CERNVMswgrowth}
\end{figure}

Looking ahead, new challenges arise for the preservation of the LHC experiment software of current and future runs.
Compared to Run 1, the software complexity rose substantially, especially with respect to external, non-HEP software such as mathematical and ML libraries and scientific Python tools~(Fig.~\ref{fig:CERNVMswgrowth}).
Growing software stacks can be mitigated by strengthening the role of software librarians and generally a disciplined approach to software dependency management.
Secondly, the last few years saw a rise of greater architectural variety compared to the previous decades which were dominated by the Intel X86\_64 architecture.
Today, there is a mix of Intel, ARM, and POWER CPUs, accompanied by GPUs and (rarely) FPGAs.
When transitioning a software stack into a frozen state, it will likely be necessary to streamline the available binaries to one or few reference architectures.

\subsection{CERNLIB}

%\todo{TODO:how to manage retirement of external s/w packages?}

The management of the retired external software packages poses a problem for the 
rapidly developing but not optimally managed physics-related software.

One of the approaches that could be used to maintain the external software packages is to create a digital archive/library implemented on top of the popular software 
versioning platforms as \github or \gitlab\cite{Schwickerath:2023rtg}.

%\subsection{CERNLIB}
\label{DPTECH:CERNLIB}

% CERNLIB: how to manage community software releases/validation on new OS / hardware
%\todo{TODO: ref for CERNLIB, DPHEP ws with CERNLIB discussion, link to gitlab project, more words on current status?}

CERNLIB is one of the key software components that have been extensively 
re-used by many collaborations and still plays an key role 
for the preservation activities e.g.\  for LEP- but also HERA-era experiments. 
The CERNLIB libraries and tools have since their 
inception been migrated to a large set of platforms and 
also several code management systems -- still preserving most of the 
original functionality and development(commit) history. 

%\paragraph{CERNLIB content and dependencies}
%\subsubsection{CERNLIB content and dependencies}
The CERNLIB contains multiple software sub-libraries, many of which are actually 
independent software packages.
Logically CERNLIB is split into a set of sub-packages
\begin{itemize}
\item {\sc mathlib}, a library with mathematical routines
\item {\sc packlib}, general purpose library
\item {\sc kernlib}, general purpose library
\item {\sc geant}, a version of the Geant3~\cite{Brun:1987ma} simulation toolkit.
\item {\sc mclib}, which contains a set of most used MC event generators.
\item {\sc graflib}, a library with interfaces to graphical routines
\item {\sc pawlib}, Physics Analysis Workstation
%\item {\sc ...}
\end{itemize}
Each sub-package is split into smaller sub-sub-subpackages, 
very often with their own requirements and compilation options.
Depending on the host system, the CERNLIB might requre the following software
\begin{itemize}
\item C preprocessor, C99 and Fortran77  compilers
\item {\sc sed} and {\sc zip} utilities
\item {\sc CMake} and any build system -- for {sc CMake} builds
\item {\sc Imake} and {\sc make} -- for {sc Imake} builds
\item BLAS with Lapack compiled with the same Fortran compiler (optionally)
\item Freetype libraries and headers
\item X11 libraries and headers of version R6.6+
\item Motif libraries and headers
\item OpenSSL libraries (and headers for CMake builds)
\end{itemize}

%\subsubsection{CERNLIB future}
%\paragraph{CERNLIB future}
%While certain historical platforms today do not anymore play an active role, several of  the LINUX platforms are still fully operational today. 
The community of users may soon be facing the challenges of a forced migration (caused 
by hardware and OS obsolescence) and consequently validation on recent 64-bit platforms. 
Since such a validation poses significant technical and personnel effort, the idea to 
consolidate the validation activities on a single, community maintained software base 
with a small set of actively validated platforms emerged at a recent DPHEP community workshop. 
The effort in a consolidated, up-to-date issue tracking and, as much as possible, automated 
build and validation pipeline would then allow to exploit the small remaining expert effort 
for the benefit all remaining CERNLIB deployments and prepare for a period in which expertise 
in CERNLIB internals is rare or not anymore readily available. 

%\subsubsection{2022 updates}
%\paragraph{2022 updates}
\label{chap:cernlib}
After the end of active development of the CERNLIB in 2007, it was widely used by many 
collaborations and individuals. The intensive usage of CERNLIB on the constantly updated operating 
systems resulted in a development of large number of code patches. In the end of 2010 most of these patches 
were consolidated 
%by Kevin McCarthy 
and used to create Debian packages for CERNLIB.
Those patches, as well as patches from the DESY theory group, Fedora and Debian projects became 
the starting point for the updates of CERNLIB in 2022. All the updates were 
implemented in the newly created CERNLIB {\sc git} repository hosted by CERN and later extended by various improvements by the 
CERNLIB/DPHEP initiative.
The repository is a merger of previously separate repositories with CERNLIB sub-libraries and contains all the 
development history of CERNLIB since 1990s.
The code is also supplemented with a continuous integration system (CI) that performs
builds of CERNLIB on selected platforms, namely older Scientific Linux 4, 5, 6; CentOS 7, 8, 9; Alma 8, 9; 
Fedora 35; in 32- and 64-bit variations, as well as Ubuntu 18, 20, 22; in 64-bit.\footnote{Upstream Canonical dropped certain 32bit development libraries, therefore for Ubuntu only 64-bit variations are available.} While some of the images used in CI
are stock images from the vendors, the older Scientific Linux 4, 5, and 6 container images are maintained under the DPHEP cover.

While CERNLIB contains codes written in Fortran and C, there are little worries about future 
ability to compile the updated CERNLIB code. However, the native CERNLIB build system, 
{\sc Imake}~\cite{imake}, is relatively outdated and can potentially disappear in the nearest future.
Therefore, while the {\sc Imake} scripts are still maintained in the CERNLIB, 
significant efforts were put recently into a re-implementation of the build system 
of CERNLIB in {\sc CMake}~\cite{cmake}. {\sc CMake} is an open-source, free cross-platform build 
tool with a history for more than 20 years. As of 2022 {\sc CMake} is de-facto the 
standard for software written in C/C++/Fortran. With the {\sc CMake}-based build system it 
became possible to perform checks of CERNLIB code with many more compilers on multiple historical and
 modern platforms. This approach is a carbon copy of the techniques used to port the JADE 
 software, see Sec.~\ref{experiment:JADE}.
As a result, the updated CERNLIB version can be compiled on RHEL4+ i686, x86\_64, ppc64 
systems and modern 64bit MacOSX using GNU 3+, CLang, Intel and NVidia compiler collections. 
However, only certain combinations of the compilers and systems are  fully backed by the 
DPHEP and included in the CI.
% We could mention here that the project was presented at ACAT 2022 in Bari, proceedings to be done
%\subsubsection{Distribution}

%\paragraph{Distribution}
The CERNLIB code including all the above improvements is available in \gitlab at CERN\footnote{\url{https://gitlab.cern.ch/DPHEP/cernlib/cernlib}}
and should serve as a drop-in replacement for the CERNLIB 2006.
The updated CERNLIB can be also packaged into RPM package.
While no further development of CERNLIB is foreseen, we encourage the current and future users to
give a feedback on the updates and to submit their suggestions, bugreports and bugfixes.

%containers versus migration?
%usage of opensource solutions from industry, e.g. build services as OpenBuild Service from 
%SuSe \href{https://build.opensuse.org/}{https://build.opensuse.org/}
%COPR \href{https://copr.fedorainfracloud.org/}{https://copr.fedorainfracloud.org/}
%CBS from CentOS \href{https://cbs.centos.org}{https://cbs.centos.org}

%\newpage
\section{DPHEP: the way forward}
DPHEP is an unique forum ensuring continuity, expertise exchange and knowledge preservation within a field where the usual HEP collaborative models are modified to deal with a rather uncommon situation: obtain new scientific results at constant level of complexity with reduced resources and therefore refactored analysis environments. With an overview of over ten years, one can observe that the concepts, the methodologies, the technologies and the collaborative configurations required to preserve data at long term require a significant amount of innovation.  While DPHEP focused so far on collider experiments, it is obvious that other collaborations -- in particle physics, astroparticles, cosmology, nuclear physics etc. -- face similar problems in preserving large and complex data sets. An increased synergy within and beyond the HEP community remains possible and desirable. The links with the industrial and economical worlds are an interesting perspective as well, still to be explored. 

Moreover, data preservation is one of the multiple possible actions towards improving the current and future computing systems in HEP and beyond. As it becomes more and more obvious from the case studies presented above, the data preservation activity is not only performing research on existing unique data sets, it is also shaping the future, in coherence with the open science policies and methodologies. Furthermore, there is a clear potential for training the new generations and improving their technological skills. 

The DPHEP Collaboration commits to drive a culture of open sharing of knowledge, data, software, algorithms, infrastructures and other research resources. The objectives for the next period are to:
\begin{enumerate}
\item Improve the awareness and stimulate improvements of data preservation:
\begin{itemize}
\item Explore and document the aspects related to data preservation: scientific motivation, organisation options, technologies, standards, outreach and education.
\item Attract new collaborators, enlarge the community, organise workshops, issue regular Global Reports.
\item Ensure links to other communities.
\end{itemize}
\item Reinforce and support the ongoing laboratory-based projects, encourage inter-laboratory cooperation and strengthen the links to other projects in different host institutions (external computing centers, Universities etc.).
\begin{itemize}
\item Keep alive data sets that (can) still produce science and keep track of parked data sets.
\end{itemize}
\item Support/develop the DP aspects for future experiments and encourage the transfer of knowledge. 
\item Encourage open data and open science as a way to preserve data and knowledge.
\end{enumerate}

%G. J. Peter Elmer to Everyone (3rd DPHEP collaboration meeting, zoom, June 23rd, 2021, 5:00 PM): 
%As our nations and communities start to recover from the pandemic and build resilience for future shocks, we will continue to work with our research and business communities to remove barriers to the open and rapid sharing of knowledge, data and tools, to the greatest extent possible. 

Although the general awareness of DP in HEP has largely increased in the past decade, and lots of activities
and successes are reported, there is still a lack of coherence between the different
experiments and projects. Transverse activities are still exceptions and cover only a minority of the existing data sets. Moreover explicit support 
of DP by some labs and funding agencies should be at least maintained and probably increase.

Data preservation is one of the building blocks of the HEP scientific outcome and the DPHEP Collaboration intends to stimulate and support it. 
%Recognising the importance of research security in particular in cutting-edge fields, and to promote open science and increase open, safe and transparent dissemination of science to citizens, and to strive to minimise technology-related risk."

\begin{comment}
    
In the specific actions:

It talks about a commitment to fostering research data management and sharing through policy, infrastructure and services, and exploring incentives ``including research assessment to drive a culture of rapid sharing of knowledge, data, software, code and other research resources."

Common solutions

Potential future development: common HEP metadata language / onthologie / data description scheme

\subsection{Sustainable models for DP?}

\subsection{Data Preservation and  Open Access?}

\subsection{What future for the DPHEP collaboration?}

What will be the impact of an increasing time gap between experiment and superseding experiment? 

Is more than one PhD-to-retirement cycle possible/proven/feasible/needed?

\subsection{What do we have to offer to the global open access scientific community?}

open data @cern, zenodo, project in Germany (get title from Achim’s talk)

\end{comment}

%%%%% acknowledgements
%\newenvironment{acknowledgement}{\relax}{\relax}
%\begin{acknowledgement}
\section*{Acknowledgements}
The authors would like to thank the respective funding agencies for support in pursuing the topic. In particular, the endorsing of DPHEP as an ICFA panel has been and continues to be an essential base of sustainable international collaboration in the area of data preservation across collaborations and laboratories.  
% add specific acknowledgements here 
%\input{acknowledgements.tex}
%\end{acknowledgement}

% BibTeX users please use
%\newpage
%%%%%%%% Bibliography 
\bibliographystyle{utphys}   % Remember we use title in the biblio
\bibliography{bibliography}
%\input {bibliography.tex}  

%%%%%%%%%%%%%%%%%%%%%%%%%%%%%%%%
% Appendices: yours (if any) 
%%%%%%%%%%%%%%%%%%%%%%%%%%%%%%%%
\newpage
\appendix
\section{Glossary}
\bigskip 
\textbf{AFS} :   Andrew File System, a distributed file system which uses a set of trusted servers to present a homogeneous, location-transparent file name space \url{https://www.openafs.org/}. \\ 
\textbf{ALCM} : LHCb Analysis Lifecycle Management.                \\ 
\textbf{ALEPH} :     Experiment at the electron-positron collider LEP at CERN, operated from 1989 to 2000 \url{https://aleph.web.cern.ch/aleph/aleph/Public.html}.                 \\ 
\textbf{ALICE} :   Detector optimized to study the collisions of nuclei at the ultra-relativistic energies provided by the LHC at CERN \url{https://alice-collaboration.web.cern.ch/}.                    \\ 
\textbf{AMI} : ATLAS Metadata Interface.                  \\ 
\textbf{AOD} : Analysis Object Data.                  \\ 
\textbf{arXiv} :  A free distribution service and an open-access archive    \url{https://arxiv.org}.          \\ 
\textbf{ASCII} :  American Standard Code for Information Interchange.                  \\ 
\textbf{ATLAS} : A general-purpose particle physics experiment at the Large Hadron Collider (LHC) at CERN \url{https://atlas.cern/}.                     \\ 
\textbf{\babar} :    Experiment operated at the PEP-II storage ring at SLAC.                  \\ 
\textbf{Belle} : Experiments (Belle I and Belle II) at KEK.           \\ 
\textbf{BEPCII} :   The Beijing Electron–Positron Collider II is a electron–positron collider.                   \\ 
\textbf{BESIII} :   Beijing Spectrometer III is the main detector for the upgraded BEPC II.                   \\ 
\textbf{BNL} :  
Brookhaven National Laboratory, NY, USA  \url{https://www.bnl.gov}.       \\ 
\textbf{CAP} : CERN Analysis Preservation Portal.                 \\ 
\textbf{CASTOR} :     The CERN Advanced Storage Manager, \url{https://castor.web.cern.ch/castor/}.      \\ 
\textbf{CC-IN2P3} : National computing centre of the French national institute IN2P3 and WLCG Tier 1. Located in Lyon, France.                 \\ 
\textbf{CDF} :  Detector and collaboration at TeVatron collider, Fermilab.                   \\ 
\textbf{CDS} :  CERN Documentation System.                   \\ 
\textbf{CERN-IT} : The information technology department at CERN.               \\ 
\textbf{CERNLIB} :  The preserved CERN Program Library is a large collection of general purpose libraries and modules maintained and offered in both source and object code form on the CERN central computers. The development is stopped since 2014 \url{https://cernlib.web.cern.ch/cernlib/}.                   \\ 
\textbf{CERNVM} : Project at CERN based on virtual machines deployment     \url{https://cernvm.cern.ch}.       \\ 
\textbf{CERN} :   European Organization for Nuclear Research (Conseil Européen pour la Recherche Nucléaire) The European Laboratory for Particle Physics, the host laboratory for the LHC collider and the ALICE, ATLAS, CMS and LHCb experiments. Located in Geneva, Switzerland.                \\ 
\textbf{CMAKE} : a framework used to control the software compilation process using simple platform and compiler independent configuration files  \url{https://cmake.org}.                 \\ 
\textbf{CMSSW} :     CMS experiment software framework.                \\ 
\textbf{CMS} :   Compact Muon Solenoid detector at LHC \url{https://cms.cern/}.                  \\ 
\textbf{CP} :     Charge Parity symmetry.                \\ 
\textbf{CSV} :   Comma-Separated Values, denotes text files that have a specific format which allows data to be saved in a table structured format.                  \\ 
\textbf{CTA} :   The CERN Tape Archive is the archival storage system for the custodial copy of all physics data at CERN. CTA is implemented as the tape back-end to EOS  \url{https://eoscta.docs.cern.ch/}.                 \\ 
\textbf{CVMFS} :  CernVM File System is CERN's software distribution service.                   \\ 
\textbf{CWL} : Common Workflow Languages.                 \\ 
\textbf{D0} :   Detector and collaboration at TeVatron collider, Fermilab operated from 1983 to 2011.       \\ 
\textbf{DAG} :  Directed Acyclic Graphs     \\ 
\textbf{dCache} :  A system for storing and retrieving huge amounts of data, distributed among a large number of heterogenous server nodes, under a single virtual filesystem tree with a variety of standard access methods \url{https://www.dcache.org/}.                   \\ 
\textbf{DELPHI} :   Experiment at the electron-positron collider LEP at CERN, operated from 1989 to 2000 \url{https://delphi-www.web.cern.ch/delphi-www/}.                    \\ 
\textbf{DESY-IT} : the information technology departement at DESY. \\ 
\textbf{DESY} :  Deutsches Elektronen Synchrotron, the host laboratory for HERA collider, operated between 1992 and 2007 with the H1, ZEUS, HERMES and HERA-b detectors. Located at Hamburg, Germany.                  \\ 
\textbf{Docker} : A software platform that allows you to build, test, and deploy applications using standardized units called containers that have everything the software needs to run including libraries, system tools, code, and runtime \url{https://www.docker.com/}.                   \\ 
\textbf{DOI} :   Digital Object Identifier.                  \\ 
\textbf{DPC} : Data Preservation Coalition. \\
\textbf{DPC-RAM} : Data Preservation Coalition Rapid Assessment Model. \\
\textbf{DPHEP} : International collaboration on data preservation in high energy physics.                 \\ 
\textbf{DST} : Data Summary Tape, a processed data format used in  collider experiments.                \\ 
\textbf{DUNE} :  The Deep Underground Neutrino Experiment, a neutrino experiment under construction. Composed of a near detector at Fermilab and a far detector at the Sanford Underground Research Facility. It will observe neutrinos produced by an intense hadron beam at Fermilab.                    \\ 
\textbf{EDM4hep} : A generic event data model for future HEP collider experiments. \\
\textbf{EIC} :  The future electron protons/ions collider at Brokhaven National Laboratory, USA.                   \\ 
\textbf{EOSC} :  European Open Science Cloud  \url{https://eosc-portal.eu/}.  \\ 
\textbf{EOS} :  EOS provides a service for storing large amounts of physics data and user files, with a focus on interactive and batch analysis \url{https://eos-web.web.cern.ch/eos-web/}. \\ 
\textbf{ESCAPE} : An Open-source Scientific Software and Service Repository  (OSSR) is a sustainable open-access repository to share scientific software and services to the science community and enable open science \url{https://projectescape.eu/}. \\
\textbf{ESD} : Event Summary Data.               \\ 
\textbf{Event} : information collected by HEP detectors for one beam crossing or collision; it contains the output of at least one collision. \\ 
\textbf{ExPaNDs} :  European Open Science Cloud (EOSC) Photon and Neutron Data Service.                  \\ 
\textbf{FAIR} :   Findable, Accessible, Interoperable, Reusable.   \\ 
\textbf{FCC} :  Future Circular Collider, the next large accelerator project at CERN.                   \\ 
\textbf{FNAL} :  Fermilab National Laboratory, the host laboratory for Tevatron collider with CDF and D0 experiments.  \\ 
\textbf{g77} :  FORTRAN 77 compiler. \\
\textbf{GAUDI} :  Software architecture and framework that can be used to facilitate the development of data processing applications for High Energy Physics experiments \url{https://gaudi.web.cern.ch/gaudi/}. \\
\textbf{\github} :    Software development framework \url{https://github.com/}.             \\ 
\textbf{\gitlab} :   An open source code repository and collaborative software development platform for large projects \url{https://about.gitlab.com/}.                   \\  
\textbf{GLANCE} :  A database retrieval mechanism for the ATLAS detector.                   \\ 
\textbf{GNU} :  A free software development framework developed in the eighties \url{https://www.gnu.org}.\\
\textbf{GridKa} : Grid Computing Centre Karlsruhe, Germany.                   \\ 
\textbf{H1} :   Experiment at electron- and positron-proton collider HERA in DESY, operated between 1992 and 2007 \url{https://h1.desy.de/}.                  \\ 
\textbf{H1oo} :  H1 Collaboration analysis software project based on Object OPriented technology in C++   \cite{Steder:2011zz}.    \\ 
\textbf{HBOOK} :  A histogram management package in PAW.        \\ 
\textbf{HEP-RC} :   The High Energy Physics Research Computing Group at the University of Victoria, Canada.                 \\ 
\textbf{HEPData} :  The Durham High-Energy Physics Database, an open-access repository for scattering data from experimental particle physics \url{https://www.hepdata.net}.         \\ 
\textbf{HEP} :   High Energy Physics.                  \\ 
\textbf{HERA} : {\bf H}adron-{\bf E}lektron-{\bf R}inganlage was a electron- and positron-proton collider at DESY in Hamburg, Germany, operated from 1992 to 2007. \\ 
\textbf{HIGZ} :  A visualisation package used in PAW \url{https://cds.cern.ch/record/281297/}.                  \\ 
\textbf{HLT} : High Level Trigger, commonly designing a last stage of selection algorithms and computing farms in HEP experiments. \\ 
\textbf{HTTP} :   Hypertext Transfer Protocol.                  \\ 
\textbf{HyperKamiokande} :    A neutrino observatory being constructed on the site of the Kamioka Observatory, near Kamioka, Japan.                   \\ 
\textbf{I/O} : Input-output.                    \\ 
\textbf{IBM AIX} : Operating system for IBM machines.           \\ 
\textbf{ICFA} :   International Committee for Future Accelerators.                  \\ 
\textbf{IHEP-CC} : Computing Center at IHEP, China.                \\ 
\textbf{IHEP} :  The Institute of High Energy Physics (IHEP), a Chinese Academy of Sciences research institute, is China’s biggest laboratory for the study of particle physics.                    \\ 
\textbf{IN2P3} :   National Institute for Particle and Nuclear Physics, French institute of the National Research Center (CNRS).                  \\ 
\textbf{INSPIRE} :   The leading information platform for High Energy Physics (HEP) literature.                  \\ 
\textbf{JADE} :   Particle detector at the PETRA particle accelerator at the German national laboratory DESY in Hamburg. It was operated from 1979 to 1986. \\ 
\textbf{JLAB} : Jefferson Lab is a DOE national laboratory focused on nuclear physics located in Newport News, VA, USA. \\
\textbf{KEKCC} : KEK central computing centre.                 \\ 
\textbf{KEK} :  The Japanese High Energy Accelerator Research Organization. It is the largest particle physics laboratory in Japan, situated in Tsukuba, Ibaraki prefecture.                   \\ 
\textbf{Key4hep} :  A project to design and provide a common set of software tools that can be used by future or even present-day high-energy physics experiments~\cite{key4hep} ~\url{https://key4hep.github.io/key4hep-doc/}.\\
\textbf{Kubernets} :  open-source system for automating deployment, scaling, and management of containerized applications \url{https://kubernetes.io/}.                  \\ 
\textbf{L3} :    One of the four experiments at LEP collider, CERN.                   \\ 
\textbf{LCG/AA} :   LHC Computing Grid Applications Area.                  \\ 
\textbf{LEP} : Large Electron–Positron Collider, operated at CERN from 1989 until 2000.  It included ALEPH, DELPHI? L3 and OPAL experiment.                   \\ 
\textbf{LHCb} :   Large Hadron Collider beauty, one of the four main detectors specialized on precise flavour measurements at LHCC, CERN.    \\ 
\textbf{LHC} :    Large Hadron Collider, the main particle accelerator (proton-proton collider) at CERN. \\ 
\textbf{m-DST} :      mini data summary tape, a reduced data format used for data analysis.              \\ 
\textbf{MC} : Monte Carlo method for collision event simulation.              \\ 
\textbf{MINERvA} :   Main Injector Neutrino ExpeRiment to study v-A interactions, a neutrino experiment using high-intensity beam to study neutrino reactions with different nuclei at Fermilab, USA.                 \\ 
\textbf{ML} : Machine Learning.                  \\ 
\textbf{MPCDF} Max-Planck Computing and Data Facility, Garching, Germany. \\
\textbf{MPP} : Max Planck Institute for Physics, Munich/Garching, Germany.                    \\ 
\textbf{NAF} : National Analysis Farm - a Computing Cluster at DESY.           \\ 
\textbf{OPAL} :  one of the four experiments at LEP collider, CERN.                   \\ 
\textbf{OPERA} : The Oscillation Project with Emulsion-tRacking Apparatus, was an  experiment for detecting tau neutrinos from muon neutrino oscillations. It was operated from 2010 to 2016 and used the CERN to Gran Sasso (CNGS) neutrino beam.                      \\ 
\textbf{OS} : Operating system.                  \\ 
\textbf{PAW} :   Physics Analysis Workstation,  an interactive, scriptable computer software tool for data analysis and graphical presentation in HEP.                 \\ 
\textbf{PEP-II} :  Electron-positron collider located at SLAC, USA. PEP-II hosted the \babar experiment, operated until 2008.                   \\ 
\textbf{PHENIX} :   Pioneering High Energy Nuclear Interaction eXperiment, is the largest of the four experiments that have taken data at the Relativistic Heavy Ion Collider RHIC, at BNL, USA            \url{https://www.phenix.bnl.gov/}.\\ 
\textbf{PIC/IFAE} : Port d’Informació Científica, a computing center at IFAE: Institut de Física d'Altes Energies, Barcelona, Spain.           \\ 
\textbf{QCD} :  Quantum Chromo-Dynamics (QCD). \\
\textbf{R2DP} : Run II Tevatron Data Preservation Project .               \\ 
\textbf{RAW} :  Acronym for unprocessed data collected at collider experiments and elsewhere (as opposed to processed data).       \\ 
\textbf{REANA} :  A reusable and reproducible research data analysis platform \url{https://reanahub.io}.                 \\ 
\textbf{RECAST} :  A framework for extending the impact of existing analyses performed by high-energy physics experiments  \url{https://recast.docs.cern.ch/}.                \\ 
\textbf{RHIC} :  Relativistic Heavy Ion Collider at BNL \url{https://www.bnl.gov/rhic/}.                \\ 
\textbf{\root} :   An open-source data analysis framework used by high energy physics and others  \url{https://root.cern}.               \\ 
\textbf{RPM} : Red Hat Package Manager (Linux OS).               \\ 
\textbf{RTA} : Real Time Analysis, online full analysis method based on intensive computing and graphical processing units.\\ 
\textbf{R\&D} :   Research and Development.                  \\ 
\textbf{Run 2} : Data taking period at LHC 2018-2022. \\
\textbf{Run 3} : Data taking period at LHC 2023-2026. \\
\textbf{SCOAP3} : partnership to convert High-Energy Physics to Open Access and to support OA publishing in these journals at no cost for authors.
\textbf{SDCC} : computing facility at BNL.                \\ 
\textbf{SL} : Scientifc Linux operating system (versions indicated as SL5, SL6 etc.). \\
\textbf{SLAC} : Stanford Linear Accelerator Center, a national laboratory  
in Menlo Park, California, USA.                    \\ 
\textbf{STFC} : UK Science and Technology  Facilities Council.              \\ 
\textbf{SuperKEKB} : Electron-positron collider at KEK.                \\ 
\textbf{TeVatron} :  Proton-antiproton collider at Fermilab, USA. Operated between 1983-2011 with two detectors and collaborations, CDF and D0.                   \\ 
\textbf{TTree} : \root data structure.                  \\ 
\textbf{VM} : Virtual Machines.                   \\ 
\textbf{WLCG} :   Worldwide LHC Computing Grid, the international common computing infrastructure and collaboration for LHC experiments.                \\ 
\textbf{XRootD} :   A framework for high performance, scalable and fault tolerant access to data repositories \url{https://xrootd.slac.stanford.edu/}.                  \\ 
\textbf{ZEBRA} : Data structure manager in PAW.                   \\ 
\textbf{Zenodo} :  Open access data and publication project based on OpenAire project at CERN \url{https://www.zenodo.org/}.                \\ 
\textbf{ZEUS} :   Experiment at electron- and positron-proton collider HERA in DESY, operated between 1992 and 2007 \url{https://www-zeus.desy.de/}.                   \\ 

%
%\input{} % put your appendices here (if any)
%

%%%%% Authorlist 
%\section{The DPHEP Collaboration}
%\label{app:collab}
%\input{DPHEP_authors.tex}  
%\end{document}
\newpage

\end{document}